\documentclass[10pt,twocolumn,letterpaper]{IEEEtran}
\usepackage{url}
\usepackage{xcolor}
\usepackage{array}
\usepackage{makecell}
\usepackage[usenames,dvipsnames]{pstricks}
\usepackage{epsfig}
\usepackage{pst-grad} 
\usepackage{pst-plot} 
\usepackage{amsmath,epsfig}
\usepackage{amssymb}
\usepackage{mathtools}

\usepackage{enumitem}
\usepackage{psfrag}
\usepackage{color}
\usepackage{pstricks}
\usepackage{cite}
\usepackage{algorithm}
\usepackage{epsfig}
\usepackage{amsthm}
\usepackage{amssymb}
\usepackage{psfrag}
\usepackage{pstricks}
\usepackage{float}
\usepackage{stfloats}
\usepackage{wrapfig}
\usepackage{graphicx,subfigure}
\usepackage{amsmath}
\usepackage{amsfonts}
\usepackage{amssymb}
\usepackage{algorithm}
\usepackage{blindtext}
\usepackage{balance}
\usepackage{pstricks,pst-plot}
\usepackage{pst-bezier}
\usepackage{pst-math}
\usepackage{algorithm,algcompatible,amsmath}
\usepackage{hyperref}
\usepackage{float}
\usepackage{cite}
\usepackage{amsmath,amssymb,amsfonts,dsfont}
\usepackage{scalerel,stackengine}
\stackMath
\newcommand\reallywidehat[1]{%
\savestack{\tmpbox}{\stretchto{%
  \scaleto{%
    \scalerel*[\widthof{\ensuremath{#1}}]{\kern-.6pt\bigwedge\kern-.6pt}%
    {\rule[-\textheight/2]{1ex}{\textheight}}
  }{\textheight}%
}{0.5ex}}%
\stackon[1pt]{#1}{\tmpbox}%
}
\usepackage{algorithm}
\usepackage[english]{babel}
\usepackage[utf8]{inputenc}
\usepackage[noend]{algpseudocode}
\usepackage{graphicx}
\usepackage{textcomp}
\usepackage{setspace}
\usepackage [english]{babel}
\usepackage [autostyle, english = american]{csquotes}

\algnewcommand\REQUIRED{\item[\textbf{Required:}]}%
\algnewcommand\INPUT{\item[\textbf{Input:}]}%
\algnewcommand\OUTPUT{\item[\textbf{Output:}]}%

\def\bp{{\bf p}}

\def\by{{\bf y}}

\def\bc{{\bf c}}

\def\bo{{\bf o}}

\def\bs{{\bf s}}
\def\bt{{\bf t}}

\def\bx{{\bf x}}

\def\so{{\boldsymbol o}}

\def\ps{\boldsymbol{p}}

\def\rx{{\mathsf{x}}}

\def\rm{{\mathsf{m}}}

        \newtheorem{theorem}{Theorem}

        \newtheorem{definition}[theorem]{Definition}

        \newtheorem{remark}{Remark}

\usepackage[normalem]{ulem}

\newcommand{\field}[1]{\mathbb{#1}}
\newcommand{\menge}[1]{\mathcal{#1}}

\newcommand{\ve}[1]{\boldsymbol{\mathbf{#1}}}
\newcommand{\vo}{\ve{o}}
\newcommand{\vs}{\ve{s}}

\newcommand{\vp}{\ve{p}}
\newcommand{\vc}{\ve{c}}

\newcommand{\Ss}{\menge{S}}

\newcommand{\Js}{\menge{J}}

\newcommand{\Cs}{\menge{C}}
\newcommand{\Ps}{\menge{P}}
\newcommand{\Ks}{\menge{K}}

\newcommand{\R}{\field{R}}

\DeclareMathOperator*\T{T}

\title{\huge Task-Oriented Communication Design in Cyber-Physical Systems: A Survey on Theory and Applications 
\thanks{$^{1}$ Centre for Security Reliability and Trust, University of Luxembourg, L-1855 Luxembourg, Luxembourg. Emails: arsham.mostaani@uni.lu, thang.vu@uni.lu, shree.sharma@ieee.org, symeon.chatzinotas@uni.lu.}
\thanks{$^{2}$ Nokia Bell Labs, Stuttgart, Germany. Emails: qi.liao@nokia-bell-labs.com.}
}
\author{\IEEEauthorblockN{Arsham Mostaani$^{1,2}$, \IEEEmembership{Student Member,~IEEE}, Thang X. Vu$^{1}$, \IEEEmembership{Senior Member,~IEEE},\\ Shree Krishna Sharma$^{1}$, \IEEEmembership{Senior Member,~IEEE}, 
Van-Dinh Nguyen$^{1}$, \IEEEmembership{Member,~IEEE},\\ Qi Liao$^{2}$, \IEEEmembership{Member,~IEEE}, and Symeon Chatzinotas$^{1}$, \IEEEmembership{Senior Member,~IEEE}}
}
\begin{document}
\maketitle
{\color{black}\begin{abstract}
 Communication system design has been traditionally guided by task-agnostic principles, which aim at efficiently transmitting as many correct bits as possible through a given channel. However, in the era of cyber-physical systems, the effectiveness of communications is not  dictated simply by the bit rate, but most importantly by the efficient completion of the task in hand, e.g., controlling remotely a robot, automating a production line or collaboratively sensing through a drone swarm. In parallel, it is projected that by 2023, half of the worldwide network connections will be among machines rather than humans. In this context, it is crucial  to establish a new paradigm for designing communication strategies for multi-agent cyber-physical systems. This is a daunting task, since it requires a combination of principles from information, communication, control theories and computer science in order to formalize a general framework for task-oriented communication designs. In this direction, this paper reviews and structures the relevant theoretical work across a wide range of scientific communities. Subsequently, it proposes a general conceptual framework for task-oriented communication design, along with its specializations according to targeted use cases. Furthermore, it provides a survey of relevant contributions in dominant applications, such as industrial internet of things, multi-unmanned aerial vehicle (UAV) systems, autonomous vehicles, distributed learning systems, smart manufacturing plants,  5G and beyond self-organizing networks, and tactile internet. Finally, this paper also highlights the most important open research topics from both the theoretical framework and application points of view.
\end{abstract}}

%
%
%
%
%
\section{Introduction}

Traditionally,  communications system design has been  guided by task-agnostic principles, which aim at efficiently transmitting as many correct bits as possible through a given channel. The design approaches have been largely based on information and coding theories, where the former sets the upper bounds on the system capacity, whereas the latter focuses on achievable techniques to approach the bounds with infinitesimal error probabilities. Despite the abstraction level of these theories, they have been successfully extended to an impressive number of communication network topologies. In this direction, digital communications has made extraordinary leaps in terms of performance, allowing robust information transfer even in the face of adverse channel conditions. 

However, in the era of cyber-physical systems, the effectiveness of communications is not dictated simply by the throughput performance indicators (e.g., bit rate, latency, jitter, fairness etc.), but most importantly by the efficient completion of the task in hand, e.g., remotely controlling a robot, automating a production line, collaboratively sensing/communicating through a drone swarm etc. It should be noted that according to projections, by 2023, half of the worldwide network connections will be among machines rather than humans \cite{cisco2020cisco}. Machines and its components (e.g., sensors, processors, actuators) - in contrast to humans - operate based on objective quantifiable processes, which in theory can be modelled and used as side information during the communication design process.  Moreover, coordination through communication messages will be imperative in terms of achieving the completion of complex tasks with the help of multiple agents, be it integrated modules such as robots, drones, vehicles or individual components thereof. In this future cyber-physical world, it becomes critical to investigate a new paradigm for designing communications strategies for multi-agent cyber-physical systems, which can be adapted or tailored on case by case basis by analyzing jointly the nature of the targeted collaborative task objective and the constraints of the underlying communication infrastructure.

Looking back to conventional communications systems, the design of the vast majority is currently based on the source-channel separation principle, which suggests that the source information can be compressed independently of the communication channel and subsequently suitable redundancy should be added to combat the adversities of the channel itself. However, it should be noted that this principle only applies under strict conditions \cite{dobrushin1959general,sourcechannel1995verdu,popovski2020semantic}. More importantly, the source compression is based on the statistical properties of the input distributions, which do not reveal the importance/value of each sample with respect to the task completion. At the same time, current communication infrastructure largely depends on the concept of layering, which is meant to create abstractions which lead to simplified system design. However, the same abstractions create rigid interfaces that prevent the higher layers (i.e., applications/tasks) from directly affecting/adapting the lower layers of communications system designs.

Incorporating this view of the problem in the design process could be a daunting task, since it requires a combination of principles from information, communication, control theories and computer science theory. In fact, all of the aforementioned scientific communities have already realised the value of this new paradigm and have made initial steps to address it from their own point of view. Nevertheless, a common framework for task-oriented communication design is still lacking, mainly due to the following challenges:
\begin{itemize}
\item Divergence of \textit{model assumptions} e.g., communication model (channel, network topology), statistical system models (partially observable Markov decision process, independent identically distributed processes), local versus global rewards/utility functions 
\item Divergence of \textit{performance metrics}, e.g., mutual information, discounted empirical error/risk, error/cost functions within time horizons
\item Divergence of \textit{mathematical tools}, e.g., rate-distortion theory, strong-weak coordination theories, dynamic programming, successive approximation, stochastic optimization and reinforcement learning.
\end{itemize}

\subsection{Semantic and Task-Oriented Communications}
{\color{black}The design of communication system mainly involves three different levels of problems, namely, technical, semantic and effectiveness \cite{Warren1949,popovski2020semantic}. Out of these, technical problems are related to the accuracy of information transmission (which may be a finite set of discrete symbols, one or many continuous functions of time and/or space coordinates) from a transmitter to a receiver. On the other hand, semantic problems are associated with how precisely and accurately the transmitted symbols can communicate the desired meaning and involve the comparison of the interpreted meaning at the receiver with the intended meaning by the transmitter based on contents, requirements and semantics. In other words, semantic communication deals with the transfer of a concept or information content from a source to the destination without going into the details of how the message is being communicated to the receiver \cite{strinati20216g,GuangmingCM2021}. In contrast to the Shannon’s framework based technical-level communication, semantic communication can provide performance enhancement due to the fact that it can exploit the prior knowledge between source and destination in the design process \cite{kountouris2021semanticsempowered,pappas2021goal,XuewenWCM22}. However,  semantic design does not consider the implications of the use fullness of the information for the task on designing the communications \cite{pappas2021goal}. The third category of problems (i.e. effectiveness) focuses on how effectively the received information can help to accomplish the desired task/performance metric \cite{popovski2020semantic,TungJSAC21}. This design paradigm is recently defined as a goal-oriented approach in the literature \cite{strinati20216g}, which is termed as task-oriented design in this paper\footnote{We prefer the term task-oriented, because the term ``goal" is often associated with humans and sounds too ambitious for cyber-physical systems comprised of machines.}. Compared with the conventional Shannon-based technical framework, new paradigms of semantic and task-oriented communications are expected to create a paradigm shift in future communication networks in terms of enhancing effectiveness and reliability without the need of additional resources such as energy and bandwidth.}

{\color{black}Task-oriented communication enables the involved entities/agents to achieve a common goal/task and its design should focus on achieving the joint objective under task-oriented constraints and specifications by utilizing the provided resources (radio spectrum, computation, energy, etc.) and suppressing the information that is not relevant to the achievement of the goal. The effectiveness of a communication design can be achieved by defining a clear goal, therefore leading to a task-oriented design. This communication framework aims to effectively fulfill the predefined goal/objective by transmitting only the  information relevant to a particular goal rather than the all raw information that would be communicated in the Shannon’s framework based approach. In such a task-oriented design, the performance of the system can be evaluated in terms of the degree by which a particular goal is fulfilled while utilizing the available amount of resources. In contrast to semantic communications, task-oriented design also utilizes the resources and entities (computation, actuation and control devices, and network nodes) usually dealt at the technical level with the objective of enhancing the effectiveness of the predefined goal \cite{strinati20216g}. As compared to the existing works focused on semantic communications \cite{kountouris2021semanticsempowered,uysal2021semantic,SemanticTheory,pappas2021goal}, the task-oriented communications in this paper will focus on the design of cyber-physical systems, which aims to enhance the task effectiveness without going into the details of semantics. Most importantly, the fact that a communication message has new semantic information does not necessarily mean that it will be useful for the task. Take the tracking example covered by \cite{pappas2021goal}. Consider some tracking information observed at the transmitter side (by the sensor) that is not visible by the receiver side (actuator). Since they are new to the receiver, this tracking information is said to have semantic value and hence worthy of communication. In a task-oriented way of thinking, however, the designer would also take into account that how this new tracking information will make any difference in the actuator's decision. If the new tracking information will not change the decision of the actuator, it has no value to be communicated - from the task-oriented point of view, see \cite{kalfa2021towards}(Sec. 4) for further readings on the differences between task-oriented and semantic-based design of communications through the lens of graph theory.
In addition, compared with the existing layered-based designs (i.e., technical, semantic and effectiveness), the task-oriented design framework in this paper does not consider the layered approach in \cite{Warren1949} but envisioned to focus on task effectiveness-based design without explicitly semantic modeling. In this regard, this paper envisages holistic policies for multi-agent cyber-physical systems to enable the joint design of communication strategies for the underlying resource-constrained B5G/6G network infrastructures and suitable action policies towards maximizing the task-oriented reward. Consequently, other information-related semantics (e.g., Age/Value of Information) should not explicitly affect the task-oriented design framework, but they could potentially be derived as a byproduct of the information distillation policy for each inter-agent connection.}

\subsection{Technological Enablers for Task-Oriented Communications Design}

The novel design methodology of task-oriented communications design (TOCD) framework departs from the conventional layered-based design and demands suitable technological enablers. In order to meet the overall system task effectiveness of TOCD, it is essential for the communications networks to be highly specialized and adapted to distinct requirements of different tasks. We envisage that these task-specific requirements can be fulfilled by the recently developed technologies such as software defined radio (SDR) and software defined networking (SDN) \cite{SDN:Akyildiz2015}, open radio access network (O-RAN) \cite{ORAN}, 5G new radio (5GNR) numerologies \cite{5GNR}. Softwarization SDR/SDN enables highly flexible PHY and NET layer configurations dedicated to different tasks via software updates which can be executed anywhere and anytime. On the system level, O-RAN creates not only a common interface for infrastructures from various vendors but also flexible functionality splits for different application needs. In parallel, it allows drawing data from various communication blocks and utilizing them to understand and optimize the performance of current configurations. Therefore, the deployment of SDR/SDN under the O-RAN environment is expected to achieve high performance and cost-efficient network configurations. On the waveform level, 5GNR numerologies allow customized radio resource blocks to meet diverse task-specific requirements without changing the transmission protocols. Furthermore, private 5G networks \cite{5Gprivate:aijaz2020private} can deliver ultra-low latency and high bandwidth connections and enhanced security level for Industrial 4.0 applications serving very large number of network elements, e.g., smart manufacturing and autonomous vehicles. A private network dedicated to a specific cyber-physical system can be fully customized to its needs based on the TOCD framework.

Despite of the rise of private networks, a large part of the communication systems will be still operating over shared infrastructure. In this context, a crucial feature of the networks to enable efficient adoption of the TOCD is the orchestration capability to harmonize the diverse, and sometimes conflicting, task-specific requirements from different applications. Such challenges create a multi-objective (cooperative and/or competitive) game on the infrastructure resource allocation (computation, storage and communication bandwidth). With the recent advances in virtualization technology, network function virtualization (NFV) \cite{NFV:Zhou2015} and network slicing \cite{NS:Zhou2016} play a key role for coexistence and infrastructure sharing to provide distinct task-oriented network configuration profiles. In cyper-physical systems, the tasks' requirements can significantly vary on both network resource (computation, storage) and communication resource (power, bandwidth). For example, a conventional eMBB applications requires a large bandwidth with tolerable latency, while an smart manufacturing application requires moderate data rate and extremely high reliable and low latency connections. In such cases, RAN slicing \cite{FlexRAN:2016} has full potential to accommodate these communications challenges. Nevertheless, we believe that the diverse requirements of the task-oriented framework can be efficiently tackled by carefully tuning of the aforementioned technological enablers without implementing additional overarching communication layers.
 
\subsection{Contributions}
{\color{black}Despite the promising studies (e.g. \cite{kalfa2021towards,strinati20216g}) and an urgent need for formalizing a general framework, there is still a large gap in the existing literature since the task-oriented communication has not been systematically reviewed. In this survey, we make a first step towards highlighting the importance of the new paradigm and surveying various approaches from the wider scientific community, which can help towards a common understanding.} Our contributions are summarized as follows:
\begin{itemize}
    \item We conduct extensive literature review from a theoretical perspective, classifying the contributions across three major communities, i.e., information/communication theory, control theory and computer science.
    \item We formulate a common conceptual task-oriented communication design (TOCD) framework to clarify and justify the selected terminology, assumptions and definitions, which will then form the basis for the general problem description for the task-oriented communications design.
    \item To validate the framework, we focus on specific use cases, where we have collected the major literature which can be studied under the prism of this new paradigm.  Properly addressed, the implications of  these topics can be far-reaching across a range of  real-world applications, such as industrial internet of things, multi-UAV systems, tactile internet, autonomous vehicles, distributed learning systems, internet of skills, smart manufacturing plants and 5G and beyond self-organizing networks.
    \item Finally, we envision a number of critical open research topics within the proposed TOCD framework and suggest potential approaches for tackling them.
\end{itemize}

\subsection{Organization}

This survey paper is structured as follows. In Section \ref{section: Prior arts}, we review the relevant literature from a theoretical perspective, classifying the contributions across information/communication, control theories and computer science. In Section~\ref{section: problem formulation}, we introduce a common conceptual framework along with the corresponding assumptions to clearly delimit the problems that are addressed within this survey. In Section \ref{section: applications}, we first focus on concrete application areas  by specifying the general framework to match the underlying system model and then structure the relevant literature. Section \ref{sec:open problems} offers a list of open research topics and challenges pertinent to task-oriented communications system design. 

\subsection{Notations}

{\color{black}
\begin{table}
\caption{Table of notations}
\centering
 \begin{tabular}{||c l ||} 
 \hline
 Symbol & Meaning \\ [0.5ex] 
 \hline\hline
 \small{$\bx(t)$} & \small{A generic random variable generated at time $t$}  \\ 
 \hline
 $\langle \bx, \by \rangle$ & Product of two vectors $\bx$ and  $\by$ \\
 \hline
  \small{$\rx(t)$} & \small{Realization of $\bx(t)$}  \\
 \hline
 \small{$\mathcal{X}$} & \small{Alphabet of \bx(t)}  \\
 \hline
 \small{$|\mathcal{X}|$} & \small{Cardinality of $\mathcal{X}$}  \\
 \hline
 \small{$p_{\bx}\big(\rx(t)\big)$} & \small{Shorthand for $\mathrm{Pr}\big(\bx(t) = \rx(t) \big)$}  \\  
 \hline
 \small{$H\big(\bx(t)\big)$} & \small{Information entropy of $\bx(t) $ (bits)}  \\  
 \hline
  \small{$I\big(\bx(t); \by(t)\big)$} & \small{Mutual information of $\bx(t)$ and $\by(t)$}  \\ [1ex]
 \hline
   \small{$\mathbb{E}_{p(\rx)}\{\bx\}$} & \small{\makecell{Expectation of the random variable $X$ over the \\ probability distribution $p(\rx)$}}  \\ [1ex]
 \hline
\end{tabular}
\label{table-notation}
\end{table}}
 The notations used throughout the paper are listed in Table~\ref{table-notation}. In general, bold font is used for matrices or scalars which are random and their realizations follow simple font. 

\section{Theoretical Concepts for Task-Oriented Communications Design} \label{section: Prior arts}
\begin{figure*}[t] 
      \centering
          \includegraphics[width=\textwidth]{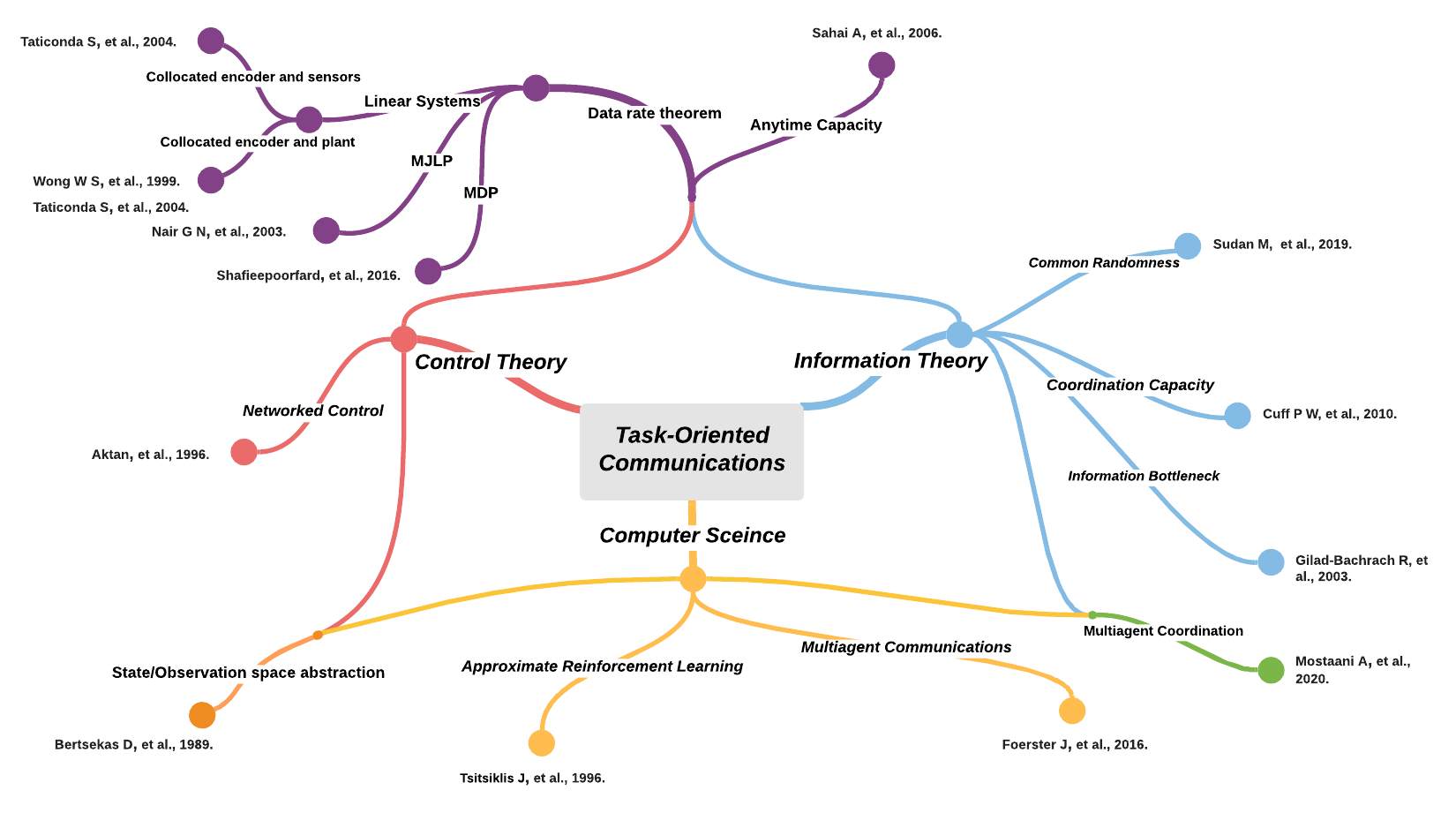}
          \vspace{-2mm}
      \caption{Main theories and selected references.}
      \label{fig: Task-based communication classification -1}
\end{figure*}

This section focuses on theoretical problems and insights which can have implications or applications on task-oriented communication design. Although the borders are often blurry, we classify contributions across three main axes: information/communication theory, computer science and control theory. Fig. \ref{fig: Task-based communication classification -1} presents a summary of the classification, main theories and selected references. 

\subsection{Information/Communication Theory}

Relevant results in Information Theory can be traced all the way back to 1971 \cite{Wyner1971Bounds}, where the state of the system as well as the observations of agents are modelled as information sources with memory and approached through the lens of rate distortion theory. The paper introduces a quantitative measure to capture the amount of information that is stored in average in the memory of the information source. It is then shown that when source coding is designed subject to any constraint on the distortion, the best achievable rate (corresponding to that distortion constraint) is lower than the achievable rate for a similar source which is memoryless. The difference between the achievable rate for a source with memory and a similar source without memory is quantified to be less than or equal to the measure of the memory of the information source. The results of the paper are general enough to apply to the sources with finite memory of any size, e.g., Markov sources with memory size of $L=1$. These results are fundamentally different
Due to the wealth of literature, this section focuses mainly on recent results and surveys, which can be used by the readers to trace back other useful references. 

The concept of common randomness \cite{sudan2019communication} is recurrent through the information-theoretic approaches relevant to our framework. Even though its applications are much wider, we focus here on the problem of generating shared uniformly distributed bits across a network of agents, using initial correlated observations as side information, complemented with interactive communication. Common or shared randomness can facilitate the execution of numerous distributed tasks, such as secret key generation, distributed computation, channel coding over arbitrary channels, synchronization, consensus, leader election etc.

\textit{Coordination Capacity}: In the information theory community, the probing of the system through different agents is often modelled as a joint distribution of actions that has to be achieved across the agents given constraints on the communication rates. In this context, the work in  \cite{cuff2010coordination} introduced the concepts of empirical and strong coordination. In this work, the concept of common randomness plays an important role and it is defined as a source of random samples which is available at multiple nodes even if there is no communication among them. Empirical coordination is tightly connected with distortion theory, since the objective is to achieve a desired joint probability distribution in a set of nodes driven by the communicated random samples generated in another set of nodes. The authors begin with toy examples of two or three nodes by examining various topologies such as the cascade and the broadcast channel. Strong coordination extends the paradigm to temporal sequences of samples. The authors in \cite{mylonakis2019empirical} introduce another variation, termed imperfect empirical coordination, aiming to bound the total variation between the joint type of actions and the desired distribution.

\textit{Anytime Capacity}: Another notable contribution in \cite{sahai2006necessity} focused on the intersection of information and control theory, by studying the concept of anytime capacity. In this case, the aim of communication is specifically targeted on stabilizing an unstable linear process (e.g., a plant control loop). While previous works have focused on erasure channels \cite{martins2004stabilization}, the work in hand addresses noisy channels. The focus is on a small-scale scalar system model with a single observer who communicated over a noisy channel with a single controller. The controller can send both control signals to the system and feedback to the observer with a one step delay. The main result is the “equivalence” between stabilization over noisy feedback channels and reliable communication over noisy channels with feedback. A key point in the model is that the decoder of the controller has to provide increasingly reliable estimates for all received past messages, as there is no side information about the message timing as required by the system. The reliability of the messages should increase sufficiently and rapidly over time to assure the stability of the system.

\begin{figure}[t] 
      \centering
          \includegraphics[width=1\columnwidth]{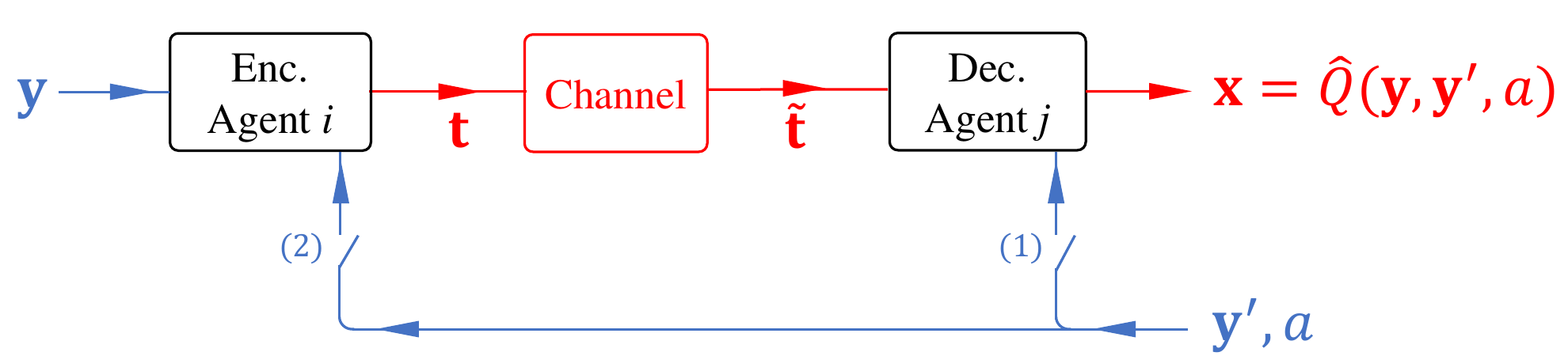}
          \vspace{-2mm}
      \caption{Distributed task based source coding.}
      \label{fig: Distributed task based source coding}
\end{figure}

\textit{Information Bottleneck}: The information bottleneck \cite{gilad2003information} is an interesting construct with hidden implications towards task-oriented communication design.  Let us consider the following formulation of three random variables:
\begin{equation}
    \begin{aligned}
    & \underset{{\bt} \in {\mathcal{T}}}{\text{max}}
    & & I (\bx ; \bt ) \\
    & \mathtt{s.t.}
    & & I(\bt; \by) < R.
    \end{aligned}
\end{equation}
The aim of the information bottleneck is to compress the information in $\by$ into $\bt$ following the rate constraint $R$, such that $\bt$ can provide the most useful information about $\bx$ in the mutual information sense.
In this context, it has been recognized that the information bottleneck problem provides a method to 
extract the information in $\by$ which is most relevant for estimating or approximating $\bx$ \cite{amjad2019learning}. 
 Now, let us consider a toy example of sequential decision making for two agents, where one receives an observation $\by$ that has to be, at least partly, communicated to the other one in order to maximize the expected cumulative reward. The random variable $\bx $ can be seen as the expected cumulative reward of the system given action $a$ and the current state of the system, where the state of the two agents is jointly defined by $\langle \by, \by' \rangle$. The parameter $\bt$ is the communication message that the first agent is about to transmit based on its observation $\by$ to facilitate the control decision $a$ by the agent $j$. Accordingly, by solving a (conditional) information bottleneck problem, at the side of encoder, we can optimize the communication of observations $\by$ of agent $i$, by compressing them to $\bt$ while ensuring that the compressed communication message $\bt$ has yet the maximum possible information about the conditional expected return $\bx$. Fig. \ref{fig: Distributed task based source coding} illustrates the setting of this problem. When switch (1) and (2) are both off the problem reduces to a rate-distortion problem with memory at encoder of the agent $i$. The work done in \cite{harremoes2007information} was first to show the connection of the information bottleneck problem and rate-distortion with logarithmic loss. To stay consistent with the framework that will be proposed in section \ref{section: problem formulation}, we assume the switch (1) to be always on. Accordingly, conditioned on the switch (2) being on, the coding in the encoder of agent $j$ should be done such that it maximizes the conditional mutual information $I\big( \bx; \bt | \by' \big)$. A solution to the conditional information bottleneck problem was provided by  \cite{gondek2003conditional}. \\

\textit{Multi-agent coordination under imperfect observations}: The contribution in \cite{larrousse2018coordination} and
\cite{lasaulce2018information} lies in the intersection of information and communication theory. The focus is on multi-agent systems that have to coordinate to maximize their long-term utility functions based on system observations impaired by an i.i.d. process. This particular example focuses on distributed power control and the system state contains the channel gains, which are partially and imperfectly known to each agent. The authors formalized the optimal problem and provided an achievable solution based on sequential best-response dynamics.

\subsection{Control Theory} \label{subsection: art - control theory}

After the advent of the networked control systems \cite{overstreet1999internetbased, aktan1996distance}, the pioneering works proposed in \cite{30,tatikonda2004control} was successful in developing a cohesive data rate theorem. The paper studies the stability of a control system under a rate-limited but reliable communication channel. It is assumed that the rate of communication is time invariable, the state process is a discrete random variable, and the disturbances that the system is subject to are bounded. 
They show that the necessary conditions on the rate of the communication channel is independent of the algorithms used for encoding and decoding the communication messages. It has also been shown that these necessary conditions on the channel rate are independent from the information patterns of the networked control system.

There has been a plethora of subsequent works afterwards to extend this setting, to more realistic scenarios \cite{nair2004stabilizability,martins2006feedback,minero2007data,31,33}. The authors of \cite{nair2004stabilizability} extended this framework to cases where the support of system disturbances is unbounded. Martins et al. accommodated time-varying rates for the communication channel in the framework, however, their results were limited to first order linear systems with bounded disturbances \cite{martins2006feedback}. The work in \cite{minero2007data} widened the applicability of the data rate theorem to finite dimensional linear systems with unbounded disturbance and time varying rate for the communication channel. In contrast to the previous works \cite{nair2004stabilizability,martins2006feedback,minero2007data}, Liu et al. in \cite{poor2020latency} considered the joint effect of all the two parameters of latency and data rate by finding a region of stability that indicates the necessary and sufficient values of data rate as well as latency. Huang also considered the joint effect of all the three parameters of latency, reliability and data rate in the control system \cite{huang2020wireless}. It was shown by Kostina that for a fully observable linear system, a lower bound for the rate-cost function can be computed even when the system disturbances are not generated by a Gaussian process \cite{kostina2019rate}. The results provided by their research can even be generalized to partially observed linear systems, when the observation noise is Gaussian. The rate-cost function is the minimum required bit rate that can guarantee the system state to be upper-bounded by a certain value.

Interested readers can find more details about the solutions proposed for the control of a linear system over a communication network in the following surveys and books \cite{matveev2009estimation}. While most of the works discussed above consider the state process to be generated by a linear system with added Gaussian system/measurement noise, fewer works have targeted the state processes which are generated by Markov jump linear processes (MJLP) \cite{nair2003infimum, song2016disturbance, zhenglong2016stabilization, zhang2009stabilization} or Markov Decision Process \cite{borkar2001optimal,Bertsekas1989AdaptiveAg,bertsekas2019biased}.

State aggregation for dynamic programming has been studied for long among the communities of control theory and operations research \cite{puterman1978modified, puterman1982action, voelkel2020aggregation, Bertsekas1989AdaptiveAg,38}. Successive convex approximation (SCA) is one of the main tools leveraged to form a trade-off between the accuracy of the solution of the dynamic programming problem and its computational complexity \cite{chatelin1982acceleration, puterman1978modified}. Later, adaptive algorithms for state aggregation were proposed which could recompose the aggregation of states during the iterations \cite{Bertsekas1989AdaptiveAg}. The major driving force for researchers, in the past, to work on this problem has been to enhance the computational efficiency of the algorithms. However, the algorithms that they have made available, can now be used to efficiently represent the state of environment while minimizing the degradation of the objective function. This area is also well explored by the community of computer science and is addressed in the next subsection.

Shafieepoorfard and Raginsky have studied the problem of controlling a Markov Decision Process while the agent is subject to observation constraint \cite{shafieepoorfard2016rationally}. In particular, the observations of the agent here are considered to have a limited mutual information with the state of the system. Most of the literature in control theory society either treats a given medium for observations as a given constraint or considers transmissions of the sensory system as an additional cost \cite{maity2018optimal}. The work in \cite{shafieepoorfard2016rationally} together with \cite{yuksel2012optimization} can be considered among the very first few papers which solve the problem of finding a stochastic control function in conjunction with the control problem. The problem of joint design of the observation and control policy is studied in this paper in its very general form as the policies are not considered as deterministic but as stochastic functions, where the transitions of the system are also considered to be stochastic. The problem of one-shot control-communication policy optimization under mutual information constraint is first shown by authors to be a form of rate-distortion problem. Armed with this analytical result, authors consequently use the Bellman equations to reformulated the core problem as a one-shot control policy optimization under mutual information constraint. This particular way that the infinite horizon control problem under observation constraint is formulated by \cite{shafieepoorfard2016rationally} was first introduced by Sims in his seminal work \cite{sims2003implications} which won him a Nobel prize.

In \cite{sims2003implications}, Sims explained the limited correlation between the behavior of economic agents and the information they have access to, observing the information through a rate limited channel. One application for this novel way of viewing/modeling economic agents is to solve the permanent income problem where there is limited information about the labour as well as the wealth, which is a dynamic programming problem with mutual information constraint on the state information of the system. In \cite{mostaani2020task} and \cite{mostaani2020state} the concept of information constrained dynamic programming is brought into the context of multi-agent coordination. Agents are limited to local observations but are allowed to communicate through rate-limited channels. Therein, the authors have developed a state aggregation algorithm which enables each agent to compress its generated communication message while maintaining their performance in the collaborative task. Similarly, in \cite{mostaani2019Learning}, authors introduced task-based joint source channel coding to solve the problem of multiagent coordination over noisy communication channels. 
To understand how these works are relevant to our framework, it should be noticed that, in fact, the information constraint on the observation of the agents in the aforementioned works can be translated as the limitation of the communication channel between the observer of the environment and the controller.

\subsection{Computer Science}
Function approximation has played an essential role in the reinforcement learning (RL) literature to overcome the limitations of the Q-learning method. The authors of \cite{tsitsiklis1997analysis,bertsekas1995neuro} have built the foundation of function approximation RL. Therein, the convergence of function approximations which are linearly combined from basis functions over a state space is established. This result opens up a wide range of applications of RL as it only requires a compact representation of the cost-to-go function, in which there are fewer number of parameters than states.
In \cite{riedmiller2005neural}, a neural network based on reinforcement learning, namely neural fitted Q (NFQ), was proposed. NFQ comprises of a multi-layer perceptron which is able to store and reuse the transition experience. It is shown therein that NFQ can effectively train a model-free Q-value and achieve the control policy after few communications rounds.
The authors of \cite{antos2008fitted} considered similar fitted Q-iteration applied to continuous state and continuous actions batch reinforcement learning. The goal is to achieve a good policy generalized from a sufficient number of generated trajectories. The authors have developed first finite-time bound for value-function based algorithms applied to continuous state and action problems. 
%
The authors of \cite{melo2008analysis} extended the temporal-difference learning proposed in \cite{tsitsiklis1997analysis} to stochastic control settings via convergence analysis of several variations of Q-learning under function approximation. Therein, the condition for the approximate methods to converge with probability one is identified. 
The advantages of approximate RL have been successfully demonstrated in real-world environments in Atari game in \cite{mnih2013playing,mnih2015human}. Therein, the deep Q-network was able to learn the policies directly from the pixels and the game score and outperformed all existing learning models.

A lossless compression scheme has been proposed by \cite{oliehoek2008optimal,oliehoek2007q, oliehoek2009lossless} for the collaborative tasks where the observations of the distributed decision makers are generated by a (Decentralized) partially observable Markov decision process (Dec-)POMDP. The authors suggest an optimal clustering scheme to partition the history of observations of an agent such that no loss in the team objective occurs. While the main intention of the paper is to reduce the complexity of distributed computations done at each individual decision maker, the proposed algorithm can have substantial applications in data compression when a group of collaborative decision makers intend to communicate through rate-limited channels.
Similarly, the work in \cite{mostaani2020task} also proposed k-medians clustering to find a proper abstraction of agents' observations before they communicate to other agents through a communication channel. It is shown that the problem of agents' observation abstraction in a multi-agent setting is, in fact, a generalized version of rate-distortion problem.

A general multi-agent framework has been proposed in \cite{LoweActorCritic} for mixed cooperative and competitive multi-agent environments. Moving from the common assumption in multi-agent studies, which assumes observations and policies of all agents are available at every agent, this work proposed to learn the approximation of other agents' policies via the maximization of the log probability of the agent. In order to maintain the robustness of the of proposed policy, the agents employ policy ensembles that are trained for a wide range of policies. Nevertheless, the developed framework allows to evaluate the impact of the communication among the agents on the learning policy by limiting the observed information from other agents. 

Efficient communication strategies have been studied in \cite{LASG1:2020} and \cite{LASG2:2020} to save the communications in distributed learning systems. Therein, the authors have proposed a so-called LASG (Lazily aggregated stochastic gradients), which adaptively determines when to communicate. The main idea of LASG is based on the new communication rule that determines the informative content of the gradients after each round based on the difference between the fresh and staled gradients. By properly implementing the rules, both downlink and uplink loads can be reduced, since only nodes with certain informative update communicate. It was shown that the proposed LASG therein achieves similar convergence as the conventional stochastic gradient descent (SGD) and significantly reduces the communications load.

The authors of \cite{pmlr-v48-abel16} proposed  a framework for states abstraction via aggregating original states into abstract states to reduce state and action spaces. Based on the unified state abstraction framework in \cite{Li06:StateAbstractionMDP}, the authors therein proposed and proved the existence of four types of approximate abstractions that guarantee the gap to the reward using the true state value function. The developed framework therein was tested in five different problems which shows significant reduction in abstract state space and suboptimal value function. 
In \cite{nachum2018near}, a representation learning scheme was proposed for hierarchical reinforcement learning in which a high-level controller learns to communicate the goal representation to the lower-level policy that is trained to achieve the goals. It was shown that a good choice of representation learning policy leads to a bounded suboptimality. Although developed for the single-agent system, these frameworks are useful to reduce the communication message in multi-agent networks.

\section{Task-Oriented Communication Framework and Scope} \label{section: problem formulation}
As surveyed in the preceding section, the task-oriented communication design has been investigated through many different viewpoints. There can be found many direct/indirect task-oriented design schemes of the communications system in the literature of computer science, information/communication theory and control theory (e.g. \cite{ioannou1988theory,narendra1978direct}). While the theoretical tools introduced in Section II were mostly focused on indirect schemes, the succeeding section will explore direct schemes. In contrast to indirect schemes, the direct schemes aim at guaranteeing or improving the performance of the cyber-physical system at a particular task by designing a task-tailored communication strategy. Since direct schemes are specifically designed for a particular task, they can hardly be generalized to other application scenarios. On the other hand, direct schemes can take the advantage of the available side knowledge about the particular task which are meant to facilitate. 
%

Therefore, we believe that having a unified problem framework would allow us to find commonalities among various task-oriented communication problems, use the available methods in the literature to solve a wider range of task-oriented communication problems and distinguish the differences between them. 
In this section, we formulate a framework, which is sufficiently generic to capture various examples of task-oriented communication design, as will be shown later in Section~\ref{section: applications}. Furthermore, the proposed framework will establish a common terminology, clarify the underlying assumptions and delimit the targeted problem set. 
%
%
\begin{figure}
      \centering
          \includegraphics[width=0.49\textwidth]{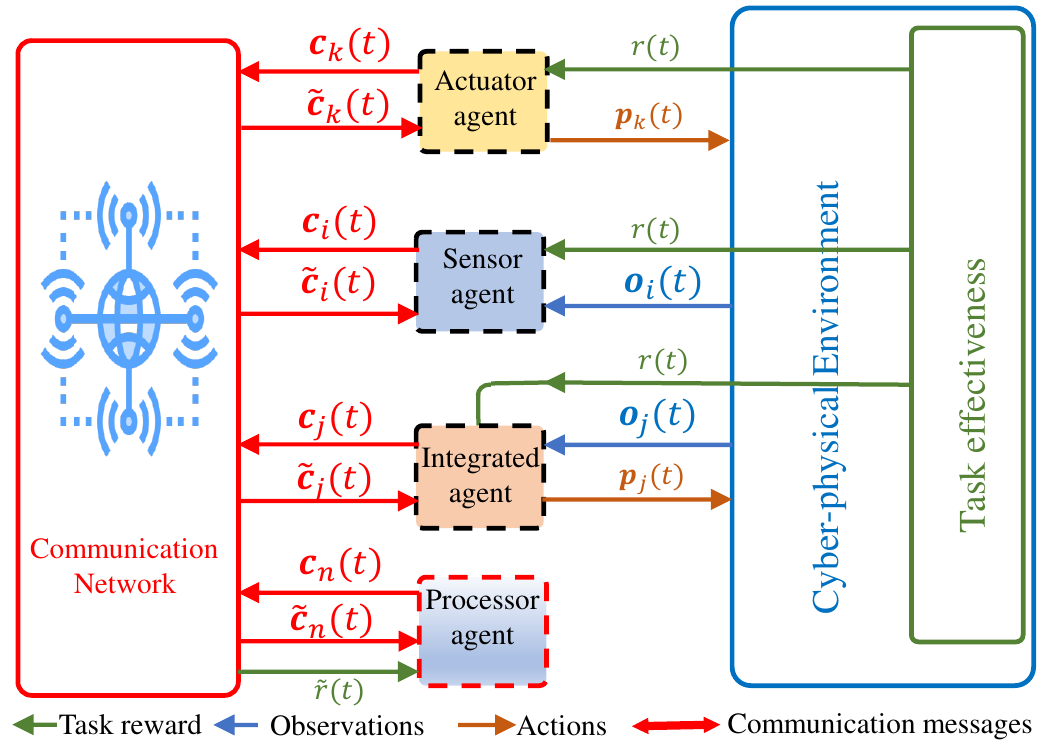}
          \vspace{-0.5cm}
      \caption{Proposed task-oriented communication design framework for cyber-physical systems. There are four types of agents with different levels of interaction with the environment.}
      \label{fig: general framework}
\end{figure}

\subsection{TOCD Overview and Agent Types}
The proposed TOCD framework shown in Fig.~\ref{fig: general framework} targets a proactive design approach to enhance the task effectiveness of cyber-physical systems, which are captured via three major components: $i$) the environment module, $ii$) the multi-agent module, and $iii$) the communication module. The environment is a core component that defines a set of parameters determining the environment state, which is controlled by the agents’ actions and is in turn translated into task effectiveness levels. The multi-agent module includes a number of agents with different levels of interaction with the environment module in terms of observations and probing actions. Finally, the communication module dictates the communication capabilities (network topology, medium access type, etc.) and constraints (rate, power, energy, interference, codeword length etc.) which dictate the nature of the inter-agent information exchange. The TOCD aims at optimizing the multi-agent module by jointly designing its communication strategies and action policies, using as input 1) the task effectiveness values from the environment module; 2) the capabilities of the multi-agent module; and 3) the constraints of the communication module. 

The TOCD framework classifies the agents into four main types: sensors, actuators, processors and integrated agents, as depicted in Fig.~\ref{fig: general framework}. The sensor agents directly  receive the task effectiveness signal and observe the environment states through its sensory measurements, which will then be sent to other agents via the communication networks. We note that the actions of sensor agents do not directly change the environment states. Examples of sensor agents include the sensors in the sensor networks or separate sensory modules in industrial plants. On the other hand, actuator agents can only have access on the environment states via communication messages with other agents, but their actions directly change the state of the environment. The integrated agents are the most complete and contain the features of both sensor and actuator agents. Therefore, they fully interact with the environment including direct observation of the system states and influence the environment via their actions. The fourth type of agent, processors, helps other agents with heavy computational or consensus tasks. Therefore, they do not directly observe or influence the environment state. In fact, the processor agent receives feedback on the current task effectiveness via the other agents through the communication network. It is worth noting that although sensor and processor agents do not directly send probing signals (actions) to the environment, their actions still have impacts on the system states by influencing other agents' actions. 

\subsection{Joint Communications and Actions Policies in TOCD}
Consider a multi-agent system with $K$ agents $\Ks:=\{1, \ldots, K\}$. A generic agent $k$ at any time slot $t$ can observe the system state through the local observation signal $\bo_k(t) = \vs_k \in \mathcal{S}_k$, where $\mathcal{S}_k$ is the set of the local system states. Then, the global system state at time step $t$ can be represented by $\vs = [\vs_1, \vs_2, \dots, \vs_K]\in\Ss$, where $\Ss:=\bigcup_{k\in\Ks}\Ss_k$. 
An agent $k$ at any time step $t$ can execute an action $\ps_k(t) $, which can affect the overall state of the system. 
Let us denote $\vp:= [\ps_1(t), \ps_2(t), \dots,\ps_K(t)] \in \mathcal{P}$ as the system action(s). 
Furthermore, let $\vs(\cdot)$ and $\vp(\cdot)$ denote the sequence of the system states and the actions over time, respectively. The overall objective is to complete an abstract task whose performance modelled by a task effectiveness function $\T(\so(\cdot), \vp(\cdot)) \in [0,1]$.
In order to make a decision on $\ps$,  we need to extract the information $\boldsymbol{c} = [\vc_1(t), \vc_2(t), \dots, \vc_K(t)] \in\Cs$ contained in $\vs$ that is relevant to the task, where $\vc_k(t)$ is the communication message sent by agent $k$ at time step $t$. The communication network between the agents is characterized via the communication operator $h:\Cs\to\widetilde{\Cs}:\vc\mapsto \tilde{\vc}$, where $\tilde{\vc} = [\tilde{\vc}_1(t), \tilde{\vc}_2(t),\dots,\tilde{\vc}_K(t)]$ and $\tilde{\vc}_k(t)$ is the message received by agent $k$ at time step $t$. We note that while $\mathcal{C}$ stands for the alphabet from which the agents can select their communication messages, the set $\tilde{\mathcal{C}}$ is the set of complex numbers as the received signal can be a complex number. 
%
%
 The task-oriented communication and action problem is defined as follows.

\begin{definition}
	The TOCD aims at jointly optimizing the \textbf{communication policy} and \textbf{action policy} to maximize the task effectiveness $\T(\vs(\cdot), \vp(\cdot))$, defined as below.
	\begin{itemize}
		\item \textbf{Communication policy} $\pi^{(C)}:\Ss\to\Cs: \vs \mapsto \vc$. Since the extracted information from $\vs\in\Ss$ needs to be transferred to the decision maker(s) via the communication channel(s) $h:\Cs\to\tilde{\Cs}$, the communication policy $\pi^{(C)}$ may include both information distillation and source/channel coding policies.
		\item \textbf{Action policy} $\pi^{(P)}: \Ss\times \widetilde{\Cs}\to\Ps: \vs \times \tilde{\vc} \mapsto \vp$ that decides the action $\vp$, that can be either a global action in the coordinator or the distributed actions in the local agents, based on the observed system states $\vs$ and the communicated information $\tilde{\vc}$.  
	\end{itemize}  
	\label{def:TOCC}
\end{definition}

Note that the above-defined problem can be subjected to additional constraints of the communications operator $h$. 
Compared against the conventional optimization problem that searches for an optimal decision policy by directly communicating states $\pi^{\ast}:\Ss\times \Ss\to\Ps$, the TOCD has twofold advantages: 1) \textit{reduce the communication cost} because the communication policy $\pi^{(C)}$ guarantees $ H(\vc)\leq H(\vs)$, i.e., the entropy of the sent information is smaller than the raw states, and 2) \textit{reduce the complexity to optimize the decision process}, because with $H(\tilde{\vc})\leq H(\vc)\leq H(\vs)$ the input space to learn the action policy $\pi^{(P)}$ can be reduced.

%

   \begin{figure*}[!b]
	\normalsize
	\begin{center}
	\includegraphics[width=0.95\textwidth]{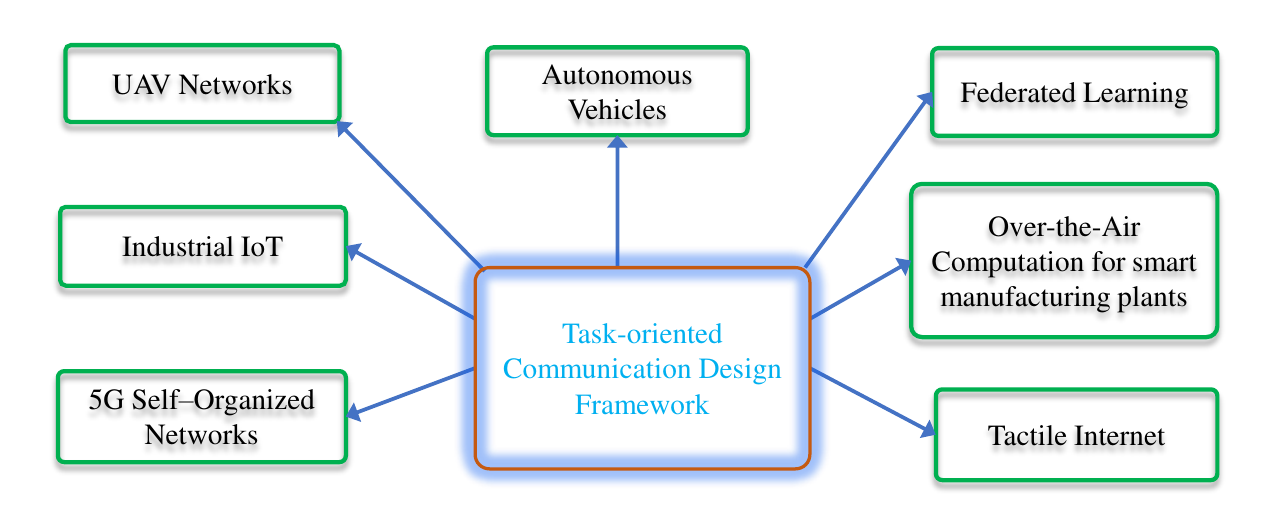}
	\caption{Application areas of the proposed TOCD framework.}
	\label{fig: applications}
\end{center}
\end{figure*}
The proposed TOCD does not assume a specific model for the state evolution of the system, although any task-oriented design of the inter-agent communication requires a consistent and unique objective function to be in place. The objective function here is the basis upon which we can measure the performance of the agents' collaborative probing of the system and it is assumed to be shown as a side information based on the targeted task. In some example scenarios, the objective function can be the accomplishment of an industrial task with enough precision in an industrial IoT framework, e.g., not missing or mistaking the target within a time horizon in a UAV object tracking framework, or having a minimal sum of errors throughout a limited time horizon in the central server of a distributed training system.
Nevertheless, the proposed TOCD sets a list of fundamental assumptions which justify the need of cooperation among the agents, as follows: 
\begin{itemize}
    \item The framework consists of at least $K\geq 2$ agents.
    
    \item There is a single, common, and consistent single-variate objective function, which is the effectiveness of the task at hand and it should not vary through the time horizon for which the problem is solved. The study for competing or non-aligned objectives for different agents is not considered by our framework.
    \item There is at least one agent with strict local observation, i.e., $ H\big( \vs \big) > H\big( \vs_k \big) \geq I\big(\vs_k; \vs  \big) > 0$ for some agent $k$, otherwise there will be no need for communication among agents, where $I(.;.)$ denotes the mutual information operator.
    \item The actions selected by agents affect both the obtained reward as well as the state process. In other words, we are less interested in the scenarios where state and action processes are independent - these scenarios usually arise in distributed estimation problems which form another rapidly growing literature \cite{sedighi2021localization, liu2021rate, kipnis2021rate, shlezinger2021deep,dommel2021joint,EhsanTSP21,ZiyangTWC21}. \footnote{Note that in these scenarios, the state-process is usually considered to be momoryless source of information which technically differentiates between the methods that are useful distributed estimation problems and task-based communication problems that are the scope of our work.}
    \item We assume that the  local/global response signal of the system is available to all agents at no cost. In case that agents have local response signals, we assume that the global response signal can be represented as a function of the local responses. 
\end{itemize}
In the next section, we will show how the proposed TOCD framework can be applied in various application domains.
\begin{remark}
	We do not assume that  the objective function is known in its analytic form by the task-oriented communication designer. Potential side information about the equations governing the evolution of the system state and the objective function can be exploited to design task-oriented communication policies using optimization/dynamic programming techniques. Nevertheless, even if we only have access to sampled data points of the objective function, one could resort to machine learning techniques for data-aided communication policy design. Machine learning techniques can also be promising to give rather generic solutions to the problem of task-oriented communication design, since they can generalize over tasks and systems that are not analytically tractable. 
\end{remark}

\section{Applications} \label{section: applications}
In this section, we demonstrate how the proposed framework captures the most popular applications of task-oriented communication networks, covering seven application areas as depicted in Fig.~\ref{fig: applications}. Furthermore, in order to understand how the literature employs theories and tools, we provide Table II which maps the application scenarios to the major task-oriented design and learning techniques. 

\begin{table*} \label{table:tools}
\caption{Applications classified by tools/techniques}
\begin{tabular}{|c|c|c|c|c|c|c|}
\hline
                                                                          & \begin{tabular}[c]{@{}c@{}}Data-rate\\ Theorem \&\\ Networked \\ Control\end{tabular}                                                                                                                                                                                          & \begin{tabular}[c]{@{}c@{}}State\\ Abstraction/\\ Approximate RL\end{tabular} & \begin{tabular}[c]{@{}c@{}}Information\\ Bottleneck\end{tabular} & \begin{tabular}[c]{@{}c@{}}Multi-agent\\  Communications\\ \& Coordination\end{tabular} & \begin{tabular}[c]{@{}c@{}}MDP and its\\ Extensions\end{tabular} & \begin{tabular}[c]{@{}c@{}}Domain-Specific\\ (non generic) \\ Methods\end{tabular} \\ \hline
                        \hline                                                   
Industrial IoT                                                            & \begin{tabular}[c]{@{}c@{}}\cite{huang2020wireless,poor2020latency}  \\ \cite{minero2009data,Petar2020NetControl,liu2019real,pezzutto2020adaptive,YuleiIoT21,NguyenTCCN18}
\end{tabular} &   -                                  
&   -
&   \cite{ZhenIoT2022}                                   &  \cite{huang2020wireless,ZhenIoT2022}             &   -                                   \\ \hline
\begin{tabular}[c]{@{}c@{}}Distributed\\ Learning\end{tabular}     
&  -
&  -
&  \cite{ullmann2020information}
&  -
&  -
&  \cite{lin2017deep,seide20141, wen2017terngrad, zhou2016dorefa, amiri2020convergence, amiri2020machine, mohri2019agnostic}                                    \\ \hline
\begin{tabular}[c]{@{}c@{}}Self-Organized\\ Networks\end{tabular} 
&  -
&  \begin{tabular}[c]{@{}c@{}} \cite{li2017user,mohajer2020mobility,xu2019load} \\ \cite{amiri2019reinforcement,mwanje2016cognitive,munoz2015load} \end{tabular}
&  -
&  \cite{li2017user}
&  -
&  -    
\\ \hline
\begin{tabular}[c]{@{}c@{}}Over the air\\ Computations\end{tabular}  
&  -
&  -
&  -
&  -
&  -
&  \cite{liu2020over,dong2020blind,frey2020over}    
\\ \hline
Tactile Internet                        &  \begin{tabular}[c]{@{}c@{}}\cite{otanez2002using,5670635,hirche2007transparent}  \end{tabular}
&  \cite{mukherjee2020leveraging}
&  -
&  -
&  \cite{assmann2021tactile}
&  \cite{shahabi2002comparison, 4374151}                                    \\ \hline
\begin{tabular}[c]{@{}c@{}}UAV/ \\Autonomous\\ Vehicle\\ Networks\end{tabular} 
&  -
& \begin{tabular}[c]{@{}c@{}} 
\cite{Cui2020,Hu2019,Wang2019a,Qie2019,Liu2019,NamIoT22,ZhengTNSE22,WeichangTITS21,TaoIoT21,HuanTETCI21,Ruilong22,LiangTMC21,ChengTWC22} \\
\cite{zhong2014reinforcement,alsamhi2020convergence}
\\
\cite{mota2021emergence,Pop2019}\end{tabular}
&  \cite{Namodeling2019,Li2019}
&  \begin{tabular}[c]{@{}c@{}}
\cite{tomashevich2018improved,Swarm:Httenrauch2019DeepRL,UAVSwarm2019,sung2020distributed}\\
\cite{alsamhi2020convergence,Loke2019} \\
\cite{mota2021emergence,Ruilong22}
\end{tabular}
& \cite{alsamhi2020convergence} 
&  -    
\\ \hline
\end{tabular}
\end{table*}

 \subsection{Industrial IoTs} \label{subsection: applications - IIoT}
\begin{figure}
	\centering         
	\subfigure[Collocated plants and sensors]{\includegraphics[width=0.47\textwidth]{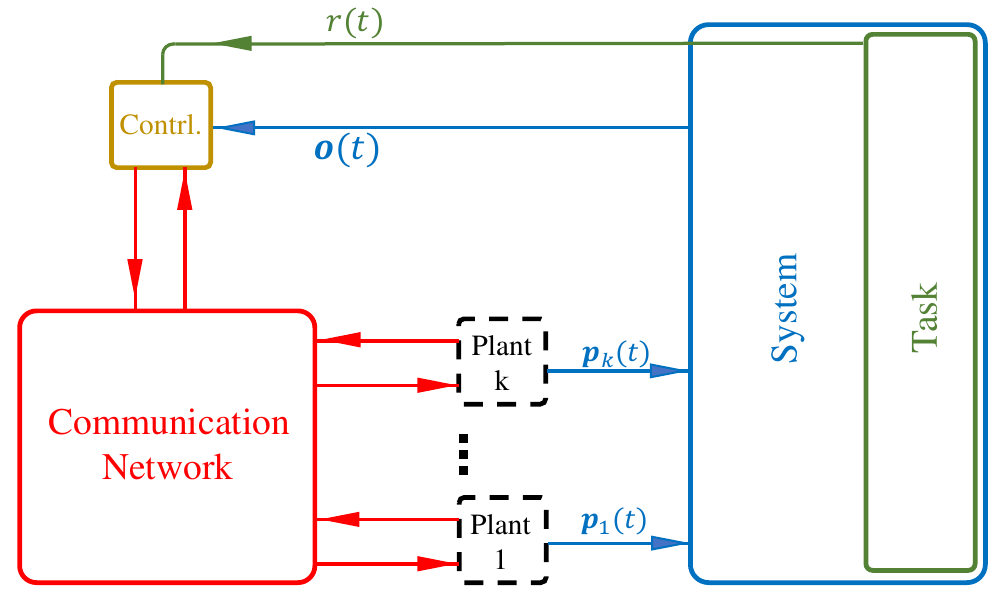}}
	\subfigure[Collocated controller and sensors]{\includegraphics[width=0.5\textwidth]{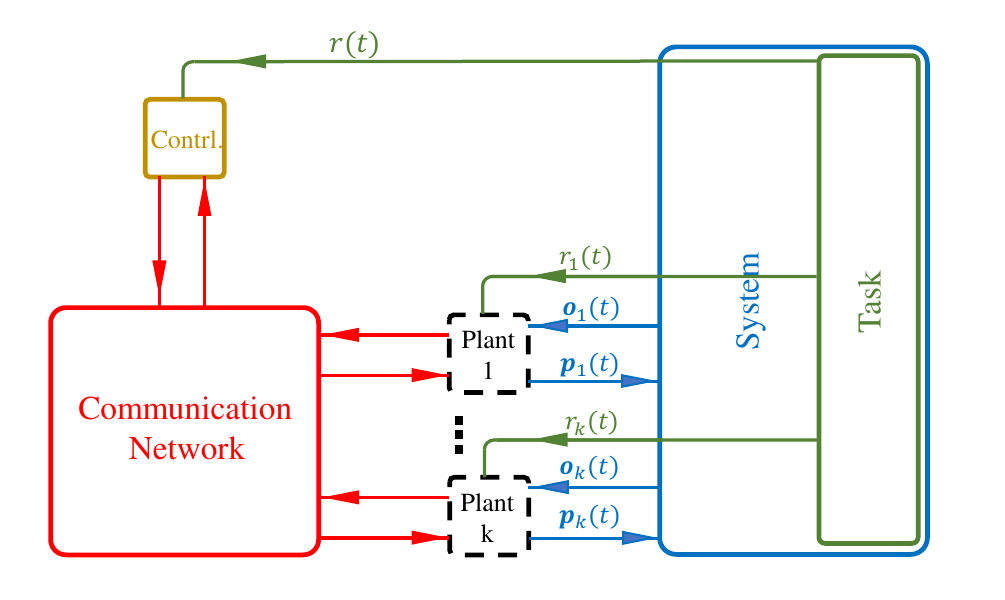}}
	\caption{An industrial internet of things problem illustrated using the Task-Based Communication framework.}
	\label{fig:IIoT}
\end{figure}

     The industrial Internet of Things (IIoT) is the generic framework that exploits the abundance of available data being generated by sensors and other devices to improve the efficiency, reliability and accuracy of an industrial manufacturing process. The availability of data generated by various devices and sensors is playing a key role here which allows each manufacturing process to be performed while having access to a much more sophisticated view of the current state of the system. Meanwhile, not every observation made at any part of the system is useful for all the stages of the manufacturing process. Given possible limitations in the rate of communications at rural areas, or the limited processing power of actuators and controllers in the manufacturing process, extracting the useful information becomes of the essence. Consider only the primary activities of a manufacturing value chain, i.e. inbound logistics, operations and outbound logistics, this huge system is comprised of many thousands of tiny elements which can generate (many) megabytes of data per second. In such a huge and complex system, communication and processing power are indeed bottlenecks of the system. Task-oriented communications would facilitate and automate the process of extracting the useful data generated by any element of the system for any controller/actuator of the system.
     The block diagram of the task-oriented communications for the industrial IoTs is depicted in Fig. \ref{fig:IIoT}. In this figure, the system response $r(t)$ is the negative of the stage cost function received after the system is probed by the plant $i$ through the signal $p_i(t)$. 
     
     In \cite{Petar2020NetControl} and \cite{liu2019real}, an IIoT system is studied, where a number of plants (e.g. chemical plants or robots) are controlled by a single controlling unit through slow/fast fading channels. The aim is to reduce the communication latency to meet the ultra low-latency requirements of industry. The authors propose a coding-free communication scheme for the central controller with the plants to optimize a shared cost function among all the plants under a power constraint at the central controller. A power control algorithm is proposed to solve the problem. The works \cite{YuleiIoT21} and \cite{NguyenTCCN18}
     showed that ML-based techniques can greatly improve the system performance in terms of the spectrum sensing and sum throughout for the complex IIoT networks.
     In \cite{Petar2020NetControl}, the global stage cost function is not locally observable by each plant. Similarly, authors in \cite{Poor2013coordinatedQ} also consider a coordination problem where the system response signal is partially observable by the agents. 
     The authors of \cite{minero2009data} considered the implications of the information rate of a channel on the system stability, where the plants are collocated with sensors and their uplink channels are lossy. The effect of time-variant information rate is considered jointly with the potentially unbounded system disturbances. In particular, the authors have analytically measured the effect of information rate of the channel on the estimation error of the system state. This allows to acquire a precise requirement on the information rate of the plant's uplink channel depending on the magnitude of the unstable mode of the system. For multidimensional linear systems, the necessary information rate of the channel is acquired based on the summation of unstable eigenvalues of the open loop. In \cite{minero2009data}, however, the joint effect of communication parameters (i.e. latency, reliability and data-rate) is not considered.


    The authors in \cite{ZhenIoT2022} developed a multi-agent RL cooperative caching framework which allows edge servers cooperatively learn the optimal caching decision. Each edge server acts as an agent to individually perform the action to predict the location and content request of IIoT devices by applying the K-order Markov chain and long short term memory (LSTM). The proposed approach is capable of both improving the cache hit
ratio and reducing content access delay.
     The trade-off between the reliability of a communication message and its latency is studied in \cite{huang2020wireless} for an industrial IoT application. 
       To study this trade-off, the impact of the age of information that is received by the plant(s) from the central controller is investigated. This is done by obtaining the value function of every possible code length through value iteration algorithm. To be able to run a value iteration algorithm, the interaction of the central controller with the whole system, including the plants, is modelled by a semi-Markov decision process. Therein, the length of error correction codes is designed considering the current state of the system rather than considering only the channel state information (CSI) of the communication network. In that sense, the error correction block of the agents are co-designed with their control policy block.
     
     In \cite{poor2020latency} the system to be controlled is linear and the uplink channel of the plant(s) is considered to be noise-free. Whereas, the downlink channel of the plants is assumed to be an AWGN feedback channel addressing the needs of low-mobility industrial IoT applications. The paper obtains a region of stability that indicates the necessary and sufficient values of communication parameters (i) information rate as well as (ii) length of error correction code blocks such that the stability of the system is insured. The paper assumes an ideal quantization and control policy to be followed across the network. Accordingly, the quantization as well as the control policy consider the history of all the communication and feedback signals exchanged between the plant and the controller. The paper also computes the average lower and upper bound of the task's cost function where the bounds are obtained as functions of communication parameters, information rate and length of error correction code blocks. The authors of \cite{pezzutto2020adaptive} solved a problem very similar to that of \cite{huang2020wireless}, where the difference is that in the former, the uplink channel of the plant(s) is considered to be noisy.

     Although the separability of the estimator of the system's state from the controller is studied in a work such as \cite{schenato2007foundations}, the separability of communication and control designs remain largely unknown. The authors in \cite{mostaani2020task} have studied the separability of control design and source coding under mild conditions. 
     Necessary data rate to guarantee a bounded cost function is obtained by \cite{nair2004stabilizability, tatikonda2004control,matveev2004problem}, where communication channels are considered to be noise-free.


\subsection{UAV Communications Networks} \label{subsection: applications - UAV}
\begin{figure}[t]
	\centering
	\includegraphics[width=\columnwidth]{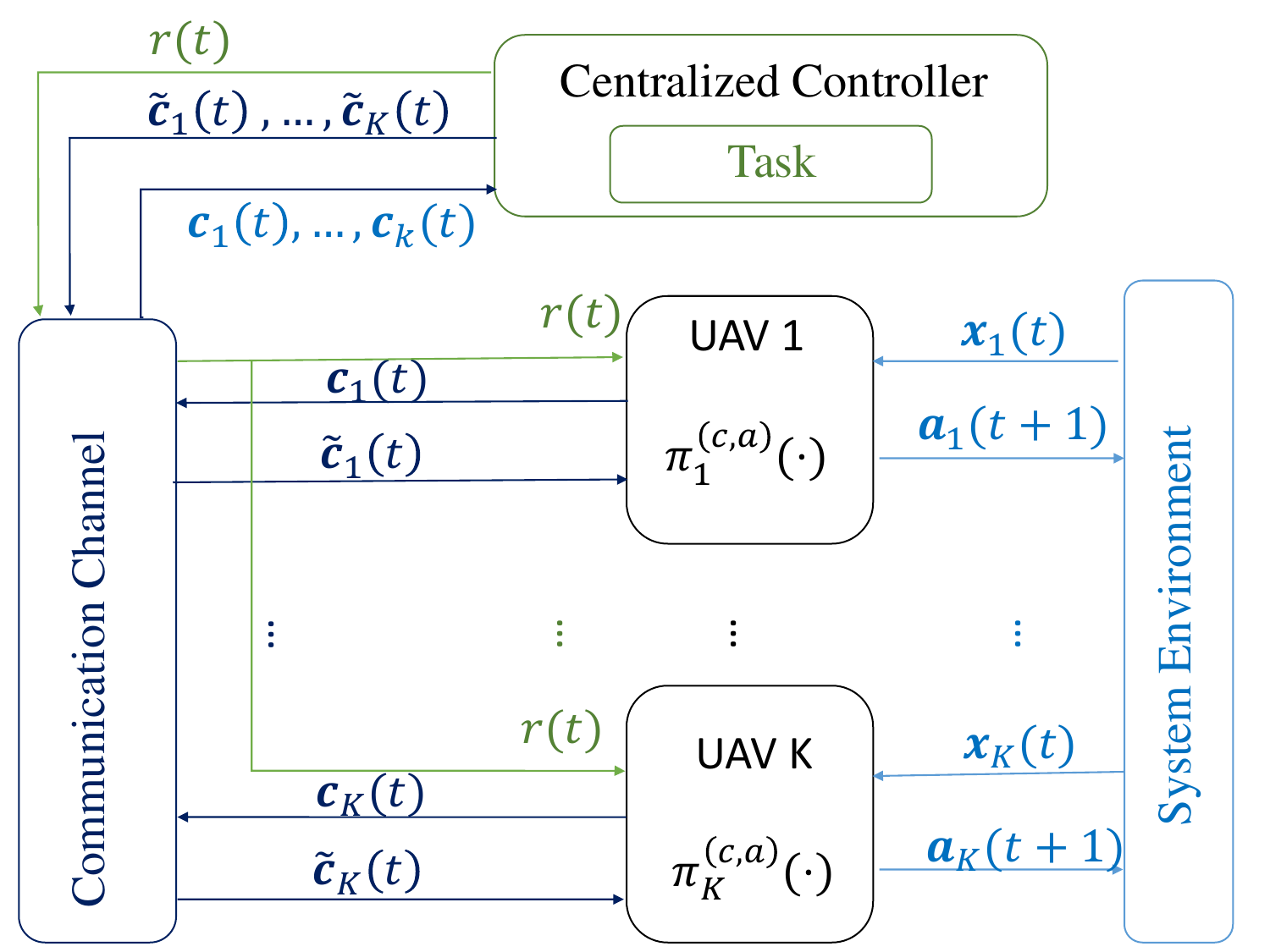}
	\caption{The proposed task-oriented communications framework applied to multi-UAV and Autonomous Vehicles networks.}
	\label{fig:UAV}
	\vspace{-0.5cm}
\end{figure}

Unmanned areal vehicles (UAV) plays an important role in the development of 5G and beyond systems due to their flexible and low-cost deployments.  UAVs can serve as a complementary application of the existing infrastructure to stand-alone service in remote areas or emergency scenarios. 
%
%
Compared to the single-UAV system, the main challenge of multi-UAV system is how to efficiently coordinate the UAVs' operations under limited communication resources. However, most of the works on multi-UAV rely on existing communications design and only focus on the UAVs' action policy \cite{Cui2020, Hu2019,Wang2019a,Qie2019,Chai2020,Faraci2020,Wu2020,Liu2020,Liu2019}. This conventional communications, which is designed for per-link performances (bit-rate maximization or packet-error rate minimization), in general is not optimal for the joint task in hand.
By using the proposed TOC framework, each UAV can jointly optimize its action policy and communication message to be exchanged with the centralized controller (or neighboring UAVs). In the $K$-UAV systems, each UAV observes the environment via its location (local state) $\boldsymbol{x}_k \in \mathcal{X}$ and received message $\tilde{\bc}_k \in \tilde{\mathcal{C}}$ from the centralized controller (or its neighbors), from which the UAV jointly determine its communication message $\boldsymbol{c}_k \in \mathcal{C}$ and next movement (action) $\boldsymbol{a}_k \in \mathcal{A}$. The performance of the collaborative task in UAV systems is then determined via a general utility function $\Phi(\boldsymbol{x},\boldsymbol{a})$, where $\boldsymbol{x} = [\boldsymbol{x}_1, \dots, \boldsymbol{x}_K]$ and $\boldsymbol{a} = [\boldsymbol{a}_1, \dots, \boldsymbol{a}_K]$ are the UAVs' locations and movements, respectively. Under the TOC design, each UAV $k$ optimizes the joint communication and action policy $\pi^{(c,a)}_k: \mathcal{X}\times \tilde{\mathcal{C}}\to \mathcal{C}_k\times\mathcal{A}$ that maps the local observation $\boldsymbol{x}_k$ and received messages $\tilde{\boldsymbol{c}}_k$ to a tuple of the local encoded message $\boldsymbol{c}_k$ and local movement $\boldsymbol{a}_k$. The block diagram of TOC framework for UAV systems is depicted in Fig.~\ref{fig:UAV}. In the following, we review the most relevant works on UAV communications systems, although most of them either assume perfect or design the communications separately from action policies.

In \cite{Hu2019}, a reinforcement learning-based sense-and-send framework was proposed for UAV networks. Therein, a BS communicates with multiple UAVs which sense data and then send it back to the BS. The objective is to maximize the sensed data sent back to the BS. To simplify the model, the authors adopt probabilistic sensing model and orthogonal sub-carriers are assumes, hence there is no interference. The channel gain is modelled as either line-of-sight (LoS) and non-LoS, depending on distance between the BS and the UAV. By modeling the system states as MDP, with three steps transmission protocol: beacon, sensing, and data transmission, a refined action space is proposed to accelerate the learning performance.
The authors of \cite{Wang2019a} proposed a Deep Q-network in multi-UAV added communications systems in which multiple UAVs serve ground users. The objective is to optimize the movements of the UAVs to maximize the system sum-rate, subject to a constraint on the minimum number of served users. Dueling DQN which uses neural network and dueling update is employed as the learning solution.
In \cite{NamIoT22}, a deep reinforcement transfer learning was developed that allows UAVs to “share” and “transfer” learning knowledge, which can reduce the learning time and  improve learning quality significantly.
In \cite{Qie2019}, the authors proposed a simultaneous target assignment and path planning based on multi-agent deep deterministic policy gradient (MADDPG) for multi-UAV system. The target is to find the shortest path for all the UAVs while avoiding collision of the task assignment. Full knowledge is assumed to be available to all UAVs during the training.
The authors of \cite{Liu2019} proposed a deep RL (DRL)-based algorithm for multi-UAV networks to optimize the 3D-deployment of the UAVs. The target is to maximize the quality of experience, defined as a weighted sum of scores of rate and delay. Q-learning and deep Q-learning is used to model the problem. However, similar to the previous work, the communication among the UAVs is not addressed. Very recently, RL-based approaches have been widely developed for the trajectory design \cite{ZhengTNSE22,WeichangTITS21,LiangTMC21}, computation-offloading scheduling \cite{TaoIoT21,HuanTETCI21}, online motion planning \cite{Ruilong22}, and energy minimization \cite{ChengTWC22}.

In \cite{Chai2020}, the authors studied the interference management of the UAV-aided cellular network, in which the UAVs and the users share the same frequency bandwidth and communicate with the BSs. Similar to \cite{zhong2014reinforcement}, the objective in \cite{Chai2020} is to maximize the average user's rate via the UAV's trajectory optimization. The problem is modeled as a non-cooperative game theory problem with full information. Thus, the communication among the UAVs is not addressed.
The authors of \cite{Wu2020} considerd the UAV-aided sense-and-send application in which the UAVs, after sensing the target, compete to access the sub-channel to send the sensed data to a mobile device. By taking into account all possible UAVs' actions, the state transition probability is derived, from which a DRL method is proposed assuming centralized architecture. 
In \cite{Liu2020}, the authors studied the UAV networks for maximizing the coverage of an area of interest, dividing into clusters. Each UAV provides service to one cluster. The goal is to design the UAVs' navigation policy to maximize the average coverage, as well as fairness among the clusters. Although modeling the problem as POMDP, it assumes unconstrained information exchange among the UAVs for free. 
The authors of \cite{tomashevich2018improved} investigated coordinated flight problem in multi-UAV systems in which the UAVs exchange their local location in order to execute given missions. Therein, an adaptive binary coding scheme is proposed based on adaptive zooming that adjusts the quantization level according the moving average of input binary signal.
In \cite{Faraci2020}, the authors proposed an RL-based framework for UAV-aided network slicing. Assuming mobile edge computing (MEC) UAV, each UAV serves a number of tasks in an area of interest. The UAV can choose to perform the jobs in its region on its own, or offload to a neighboring UAV. Assuming the job arrival follows the Switched Batch Bernoulli Process, the transition probability matrix is derived taking into account the states of the region and queues at the UAVs. Although the communication among the UAVs is considered, the communication rate is fixed and it does not actively interact with the learning process.

One of the most important research topics in UAV study is UAV swarms, a (large) group of UAVs which collaboratively perform some task. The major challenge in UAV swarms compared to UAV-assisted communication networks lies in the lack of centralized control due to its very large and highly dynamic topology. Unlike UAV-aided communications, where the interaction among the UAVs is aided by a centralized node, e.g., base station, the communication in UAV swarm is usually done via mesh networks. Due to the large topology, each UAV can obtain only partial observation of the environment and needs to cooperate with other UAVs to improve the learning process. An efficient way to leverage the collaboration in UAV swarms is to employ an interaction graph, which maintains a set of neighbor nodes for each UAV. This principle is considered in  \cite{Swarm:Httenrauch2019DeepRL} and \cite{UAVSwarm2019}, which study the impact of the communication among the UAVs to the distributed reinforcement learning task. In particular, they demonstrate via a rendezvous problem that allowing more exchanged information among neighbor UAVs can significantly accelerate the learning process and results in higher rewards. The authors of \cite{sung2020distributed} analyzed the communication impacts on multi-robot multi-target tracking problem, where each flying robot can only communicate with its neighbors within its transmission range. Therein, two learning algorithms are proposed to achieve the agreement among the robots under limited communication time. The work in \cite{alsamhi2020convergence} provided and extensive survey on the use-cases of machine learning for inter-robot communication design - including robot communications under rate-limit and time constraint to name a few. These scenarios all fall under the umbrella of TOCD for the autonomous robots/UAVs/Vehicles.

\subsection{Autonomous Vehicles}
Autonomous driving is dependent on the efficient processing of data gathered from various sensors including radar, camera, and light detection and ranging (LiDAR), and involves a complex design process to have a dependable and flexible real-time system. One important use case of autonomous driving is cooperative automated driving, in which one crucial challenge is to ensure the safety gap ($<5$ ms) between the vehicles, thus requiring stringent communication requirements in terms of latency and reliability \cite{Dressler2019cooperative}. The underlying entities should be fully coordinated with the help of suitable communication mechanisms, such as mmWave, cellular and visible light communications,  in order to ensure the full dependability of autonomous driving systems. Furthermore, in order to guarantee the reliability of information transmission via redundancy, multiple communications links could be utilized in parallel. In this regard, the proposed TOC framework can be applied to autonomous driving systems in a similar manner as in Fig.~\ref{fig:UAV}. The main differences compared with UAV systems are larger dimensions of both action and observation spaces and more stringent requirements of task effectiveness, which in consequence requires more powerful communication channels.

The transformation from manual control to fully automated driving in autonomous vehicles demands for the efficient management of control authority between the automation and human driver by avoiding human-machine conflicts. In this regard, haptic sharing control could be one promising approach to dynamically adapt the control authority between the human driver, and to suggest suitable actions while exploiting the environmental perception and the driver's state \cite{Benloucif2019}. A haptic sharing control architecture proposed in  \cite{Benloucif2019} comprises two hierarchical levels, namely tactile and operation levels, which are responsible for taking driving decisions and to provide helpful actions to track the planned trajectory of the vehicle. This approach incorporates the tactile variables such as driving activities of human driver and the control authority into the planning algorithm so that the automation can better resemble the driver's strategy for planning the vehicle trajectories. 

An important design aspect in cooperative automated driving system is the transformation of the single vehicle perception/control in self-driving vehilces to multi-vehicle perception/control, as a vehicle's perception field is dependent on the local coverage of sensors embedded in that vehicle. This requires the need of cooperative perception and manuevering \cite{During2016cooperative}, in which TI can play an important role to enable  the reliable and fast transmission of haptic information related to the driving trajectories along with sensor information via the underlying vehicular communication network. Furthermore, in cooperative adaptive cruise control (CACC) or platooning applications, which comprise several cars autonomously following the leaders, there arise stringent communication requirements in terms of reliability and update frequency to enhance the safety and traffic flow efficiency\cite{Segata2015trans}. This demands for the design of novel communication strategies for synchronized communications and dynamic adaptation of transmit power, and task-based communication could be promising in addressing these issues. 

Another important design aspect for autonomous driving is to design a human-computer interaction (HCI) system with the human-in-the-loop in a co-adaptive manner, as the complete removal of human involvement might be impractical because of various involved uncertainties including human behavior, environmental variations and user requirements. In this regard, authors in \cite{Ghosh2019} used a customizable traffic simulator, which utilizes Artificial Intelligence (AI) to predict the traffic quality at the intersection and can be used as a feedback to the human driver's decision under uncertainties. It has been demonstrated that the proposed cooperative AI-enabled decision making platform can increase the safety and average traffic delay as compared to the individual automated and human-operated traffic systems.  

Furthermore, in highly automated driving systems, it is important to enhance the trust towards automation process, and one promising approach in this direction could be to enhance the situational awareness by displaying the spatial information of close traffic objects via a vibrotactile display \cite{Sonoda2017}. Since the levels of trust in automation vary dynamically depending on the knowledge about the surrounding traffic status, the display of spatial information of nearby vehicles in a vibro-tactile display captured via a haptic stimulus can be significantly useful in designing an automated driving system. Another crucial aspect in automated driving is to ensure reliable interactions between human drivers and automated driving systems so that possible collisions due to the divergence of actions taken by human drivers and automated driving system can be avoided. The existing research works related to such interactions mainly follow the experimental approaches, which are usually expensive and time consuming. This has led to the need of efficient models for future automated driving technologies, which can predict and interpret the human driver's interaction with the automated driving system \cite{Namodeling2019}. 

In order to adapt the task-based design of autonomous vehicles in dynamic driving situations, it is crucial to gather 3D information about the road and surrounding vehicles accurately in real-time with the help of vehicular to infrastructure (V2I) or vehicle to vehicle (V2V) communications. However, most of the existing methods to capture the 3D road perception focus on a single task/aspect even if that particular aspect is not so important, thus leading to larger delay for autonomous vehicles in completing all the required tasks \cite{FUWUmultitask2020}. Although vision-based methods benefit from the use of deep learning, they suffer from the loss of 3D information. In this regard, multi-task deep learning \cite{Pop2019}  could be promising due to its potential to improve the performance of the individual tasks and to enhance the overall efficiency of the network. Pedestrian detection and estimation of time to cross the street are important issues to be addressed in the design of autonomous vehicle systems. The DL-enabled task-based design should consider a detection model, a classification model and prediction model, which deal with the localization and recognition of the pedestrians, distinguishing the pedestrian actions and estimation of pedestrian actions, respectively \cite{Pop2019}. Also, a loss function considering the learnable weight of each task can be utilized to train the underlying deep neural network in order to enhance the performance of individual tasks and to balance the loss of each task. 

Furthermore, heterogeneous service requests coming from the autonomous vehicle users comprise multiple tasks which are dependent on the availability of the limited resources. The task-based design with the success ratio of task execution as the target performance metric can be carried out at the autonomous vehicle cloud, which is connected with the underlying V2I and V2X architectures \cite{Shuotaskalloc}. Although cluster head of V2V architecture, which connect with the autonomous vehicle cloud, can allocate tasks to the vehicles for execution and can predict the task execution time, it may not provide accurate prediction of task completion time due to the limited storage and computing resources. Also, in the V2I architecture comprising vehicles and infrastructure nodes, mobility of vehicles poses challenges in designing efficient task-allocation strategies. In this regard, the efficient allocation of tasks while considering the communication node's stability and computing capability with the objective of minimizing the task completion time required to guarantee the smooth execution of requested services and to enhance the task execution success ratio is an interesting research problem \cite{Shuotaskalloc}. It should be noted that in contrast to the most existing works focused on offloading of tasks in edge computing environments, the focus of this paper is on designing task-oriented communication network with the objective of maximizing the task effectiveness metrics, i.e., the performance metrics which can maximize the tax-oriented reward. 

In cooperative automated driving applications, vehicles need to communicate not only with other vehicles but also other road-users including bicycles, motorcycles, pedestrians and road-side IoT units over the underlying 5G-V2X or short-range communications networks \cite{Loke2019}. The main crucial aspects to be considered during task-based design of these applications include how effectively vehicles coordinate with other vehicles, pedestrians and road-side units, how inter-vehicular cooperation can be utilized for better situational awareness of road-side conditions and how effective is the designed task-oriented protocol, see e.g. this work \cite{mota2021emergence}, in terms of giving collision warning at the intersections, defining mechanisms for overtaking, merging traffic and platooning in the highways, and designing policy rules for governing road traffic as well as ethical and trustworthy interactions with the central unit/cloud server. 
\subsection{Distributed Learning Systems} \label{subsection: applications - Federated Learning}
Distributed learning systems have emerged due to the growing size of training data-sets. A careful design of the communication work-load and privacy-preservation of the clients/edge devices are sought to be the enabling means in these systems \cite{shokri2015privacy,konevcny2016federated}. Lately, federated learning has received much attention as an alternative setting: a parameterized global model is trained under the coordination of a parameter server with a loose federation of participating edge devices/clients \cite{reddi2016aide, shokri2015privacy, konevcny2016federated}. Because of the major role that communication plays in all variants of distributed learning systems, it is one of the active areas of research, where direct task-oriented communication design has proven significantly efficient. This subsection details, with examples from the literature, why/how task-based communication design is helpful for distributed learning systems.

     \begin{figure}[t]
 	\centering
 	\includegraphics[width=0.50\textwidth]{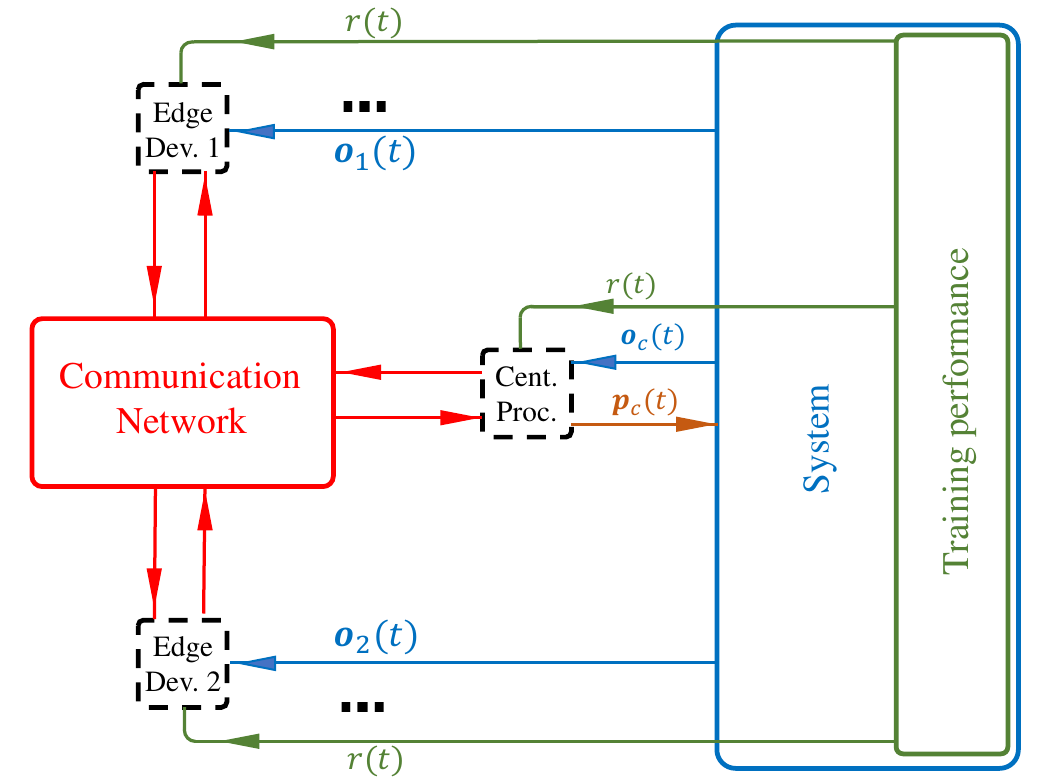}
 	\vspace{-2mm}
 	\caption{A centralized federated learning system illustrated using the task-based communication framework.}
 	\label{decentralized problem - figure}
 \end{figure}

In federated learning systems, the global model is trained using a number of different data points $\big{\{}  X_l(t) \big{\}}_{l=1 }^{n'}$ that are spread over different edge devices at time $t$. Clients are allowed to send communication messages $\bc(t)$ to the parameter server as to facilitate the convergence of the model parameters $\bp(t)$. The training data set $\big{\{}  X_{l,i}(t) \big{\}}_{l=1 }^{n'_i}$ available at client $i$ within time step $t$, together with the latest available model parameters $\bp(t-1)$ are interpreted here as the local observations of that agent/edge device in our universal framework i.e., $\bo_i(t) = \langle \big{\{}  X_{l,i}(t) \big{\}}_{l=1 }^{n'_i}, \bp(t-1)  \rangle$. We also define the state of the system to be the collection of all data points available at all edge devices, that is, the state is jointly observable by all clients. This definition of state is technically correct since (state-probation pairs $\bs(t), \bp(t)$ will still be jointly sufficient statistics for the stage cost $r\big( \bs(t), \bp(t) \big) $ and next stage state $\bs(t+1)$. Note that the stage cost $r\big( \bs(t), \bp(t) \big) $ here captures the expected loss corresponding to our training model caused at all data points $\big{\{}  X_l(t) \big{\}}_{l=1 }^{n'}$. 

In connection with our universal framework, the task of a distributed learning system is to optimize a parameterized model by solving
 \begin{equation} \label{eq: federated learning optimization problem}
 \begin{aligned}
   &  \underset{\pi^{(C)}, \pi^{(P)}}{\text{argmin}}
   & & \mathbb{E}_{\bs} \big{\{}
   \mathcal{J} \big(  \bp(t) , \hat{\bs}(t)\big)
   \big{\}}
 \end{aligned}  
 \end{equation}
 where $\mathcal{J} \big( \bp(t), \hat{\bs}(t) \big) = \sum_{t=t_0}^{t_{MAX}} r\big( \hat{\bs}(t), \bp(t) \big) $ captures the sum of losses corresponding to our training model caused at the data points $\big{\{}  X_l(t) \big{\}}_{l=1 }^{n'}$, at all times, and  $\pi^{(C)}, \pi^{(P)}$ stand for the communications and action policies defined in Section~\ref{section: problem formulation}. 

A naive strategy to carry out the communications between an edge device and the parameter server is to communicate all the local data of each edge device. The communication strategy, however, can be much more efficient if each edge device
computes an update to the current global model maintained, and only communicates the update \cite{mcmahan2017communication}. In standard federated learning, this is done by applying SGD distributively over the local data set $\big{\{}  X_{l,i}(t) \big{\}}_{l=1 }^{n_i}$ available at each edge device $i$ and communicating the average gradient of the loss function at each node $i$ and iteration $t$, to the parameter server
\begin{equation}\label{eq: federated learning - local gradients}
    \bc_i(t) = \frac{1}{n_i}  \sum_{j=1}^{n_i} \nabla_{\bp}  \mathcal{J} \big( \bp(t), \bo_{i,j}(t) \big).
\end{equation}
In fact, the large size of  data-sets and privacy of edge-devices as well as possible changes in  data-sets are the main reasons that we are not willing to communicate all the state information to the parameter server. In this sense, even the very first variants of federated learning \cite{reddi2016aide, shokri2015privacy, konevcny2016federated} introduced some form of task-based communication, where communication messages $\bc_i(t)$ of the clients to the central controller, are of much smaller size than the observations of clients and yet the task can be accomplished with no compromise on the performance. That is, the size of observations of each client is thousands of times larger than the size of gradient updates being communicated.

 However, these techniques applied, the communication between the edge devices and the parameter server is yet seen to be a major bottleneck. The authors in \cite{lin2017deep}, together with similar works \cite{seide20141, wen2017terngrad, zhou2016dorefa, amiri2020convergence, amiri2020machine, mohri2019agnostic}, introduced schemes to reduce the size of communication messages beyond the standard federated learning. These methods are shown to enhance the speed of convergence as well as to overcome communication bandwidth constraints. 
%

The authors of \cite{lin2017deep} introduced a particular coding scheme that heuristically identifies and sends the important gradients as to reduce the size of communication messages. Following the proposed scheme, the authors in \cite{lin2017deep} reported a significant task-oriented compression ratio while virtually no loss is seen in optimizing the cost function. This scheme first finds a threshold for the magnitude of the gradient vectors, above which the computed gradient of a node will be considered for communication to the master node. The gradient vectors with lower size than the threshold, however, will remain in the node and will be accumulated with the rest of gradients that will/already have computed by the same node and have not be qualified for transmission. 
While in \cite{seide20141, wen2017terngrad, zhou2016dorefa}, the communication channels are considered to be rate-limited but error-free, works done in \cite{amiri2020convergence, amiri2020machine, mohri2019agnostic} (partially) consider the effects of the physical layer features of the communication links on the problem of federated learning \cite{ullmann2020information}.

Another useful way of adopting TOCD for the purpose of distributed learning systems is to consider the value/importance of data-sets available at each client to optimize the resource management in the communication network \cite{importancereview2019wen}. As an instance, the authors in \cite{liu2020data} and \cite{molin2019scheduling} optimize user scheduling by incorporating the importance/value of the clients' data-set for the estimation taking place at the server. In particular, the work in \cite{liu2020data} introduced an indicator to capture the importance of data-set of each client, according to which scheduling of client-server communications is optimized. One unique aspect of the proposed metric, is that the scheme also considers the quality of the communication channel between a client and server to design the importance indicator. 

 One way to look at TOCD for federated learning is that a form of compression of the input data is carried out by TOCD such that we can still obtain sufficient statistics about their corresponding labels. While recent research results testify the applicability of information bottleneck with the same purpose on deep neural networks \cite{goldfeld2020information, shwartz2017opening}, very few research is done to harness the potential of information bottlenecks within the distributed learning systems \cite{ullmann2020information}.

%
%
\subsection{Over-The-Air Computation in Smart Manufacturing Plants}\label{subsec:AirComp}
In IoT networks, massive amounts of data are generated, collected, and leveraged to help complete a predefined task. For example, in smart manufacturing plants, wireless data needs to be collected from thousands of sensors. However, we are not interested in the value of each individual data source, instead, we aim to obtain the \lq\lq fusion\rq\rq \ of the information contained in all data sources, e.g., computing sums or arithmetic averages. On one hand, transferring raw measurements from a large amount of different data sources to the same data collector is not spectrum efficient, especially when the measurements can be encoded to small data packets. On the other hand, the computation of massive data in an edge device as data collector with limited computation capacity can be also challenging. 

Therefore, the technique called over-the-air computation (AirComp) has been developed to enable communication- and computation-efficient data fusion of the sensing data from large amount of the concurrent sensor transmissions. It takes into account the underlying task, such as computing a function, directly at the physical layer, by exploiting the superposition property of the wireless channel. In other words, it allows an efficient target function computation over the \lq\lq air\rq\rq. 

The AirComp is defined as follows. Consider $K$ wireless sensors, each having a measurement signal $s_k\in \R$, $k\in\Ks:=\{1, 2, \ldots, K\}$ to send. On the receiver side, we expect to derive the function of the measurements of the form 
\begin{equation}
    f(s_1, \ldots, s_k) = F\left(\sum_{k=1}^K f_k(s_k)\right).
    \label{eqn:OTA_fkn}
\end{equation}

Given the multiple access communication channel $h:\field{C}^K\to\field{C}$, AirComp aims at finding a set of {pre-processing functions} $\zeta_k:\R\to\field{C}$ and a {post-processing function} $\psi:\field{C}\to \R$ such that 
\begin{equation}
f(s_1, \ldots, s_k) = \psi\Big(h\big(\zeta_1(s_1) \ldots, \zeta_K(s_K)\big)\Big).
    \label{eqn:OTA_obj}
\end{equation}
With the pre-processing of the measurement signal, as well as the post-processing of the received channel output, we can directly obtain the desired computation of the target function in \eqref{eqn:OTA_fkn} by effectively integrating the communication and the computation policies. Moreover, in this way the receiver's computational task $f(\cdot)$ defined in \eqref{eqn:OTA_fkn} of processing $K$ signals is decomposed into $K+1$ small tasks $\left\{\zeta_1(\cdot), \ldots, \zeta_K(\cdot), \psi(\cdot)\right\}$ that can be distributed among sensors and the receiver, with each of them processing only one signal. 

\begin{figure}[t]
          \centering
              \includegraphics[width=\columnwidth]{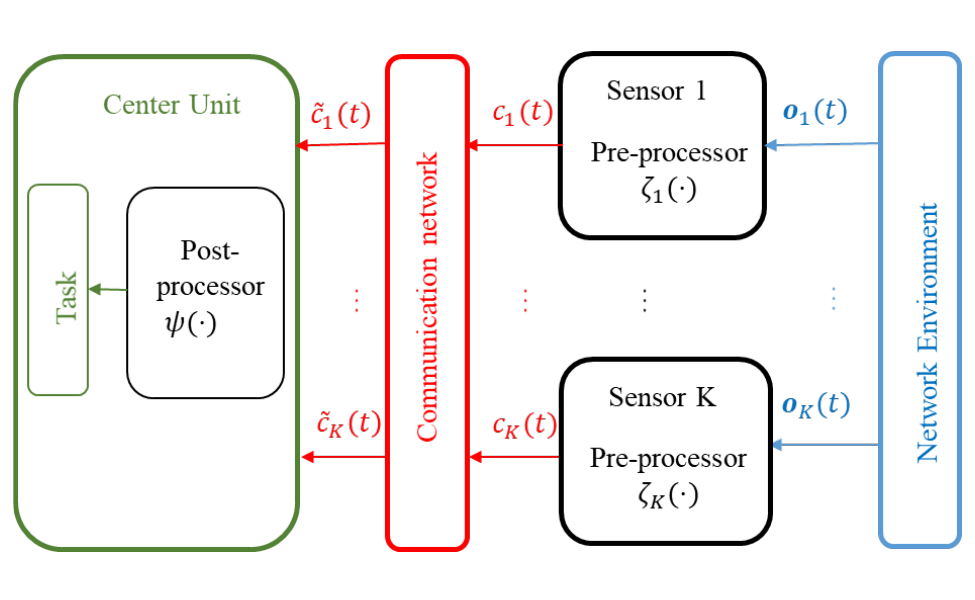}
          \caption{Over-the-air computation in smart manufacturing plants aligned with the task-oriented communications system design}
          \label{fig:OTA}
    \end{figure}

Without loss of generality, AirComp also falls into the class of task-oriented communications system design. The task is to compute $f(s_1, \ldots, s_k)$ at the receiver (e.g., a center unit) side. The problem defined above can be aligned with the task-oriented communication framework defined in Section \ref{section: problem formulation} by simply integrating the computing policy into the communication policy, i.e., the encoding (pre-processing) and decoding (post-processing) policies included in the communication policy already provide the direct action to compute the desired target function as shown in Fig. \ref{fig:OTA}.  

The authors of \cite{liu2020over} derived theoretical bounds on the mean squared error for a certain AirComp function (sum of the signals) computation in a fast-fading scenario with CSI available at the transmitter.  In \cite{dong2020blind}, the AirComp problem is tackled without the explicit knowledge of channel information but under the assumption of slow fading.  Then, in \cite{frey2020over} a AirComp scheme was proposed for distribution approximation of a larger class of functions than the previous works with theoretically proven bound over fast-fading channels that can deal with correlated fading and requires no CSI.

A popular application of AirComp computational techniques is the computation of distributed gradient descent to solve the empirical risk minimization problem for machine learning models. One example is the training of neural networks in a central mode but with distributed data reported by a large number of local agents such as sensors in the smart manufacturing plant. In \cite{amiri2020machine}, the authors proposed to use AirComp computation over wireless channels to help efficiently compute distributed stochastic gradient descent in the federated learning paradigm. The author in \cite{yang2020federated} extended the idea to channels with fading channel information at either the transmitter or receiver side. In \cite{frey2020over}, the authors showed the application of a proposed AirComp scheme to the regressor and classifiers in vertical federated learning.

%
%
\subsection{5G and Beyond Self-Organizing Networks}
Many works proposed reinforcement learning-based solutions for various self-organizing network (SON) use cases, as well as the coordination between the SON functions \cite{amiri2019reinforcement,mwanje2016cognitive,munoz2015load}. However, in these works, the selection of the network state information and the optimization of the learning function were considered as two independent processes. Expert knowledge is exploited to select the features and use them as the network state information, which may cause either insufficient information, thus poor optimization performance, or too much redundant information, thus high complexity of the learning model. 

The task-oriented communications system design can be leveraged to jointly optimize the information exaction and control optimization problems. Let us take the SON function mobility load balancing (MLB) as an example. MLB is a function where cells suffering congestion can transfer load to other cells which have spare resources, by adjusting their handover (HO) control parameters (for details of HO parameters refer to \cite{lobinger2011coordinating}). Many works have proposed to solve the multi-agent MLB problem with the following centralized model and manually selected observations by using deep reinforcement learning \cite{li2017user,mohajer2020mobility,xu2019load}. Given a set of cell sites (hereafter referred to as cells) $\Ks =\{1, 2, \ldots, K\}$, each cell can configure its HO control parameters $\vp_k\in \R^N$ and obtain a partial observation $\vo_k\in \R^M$ of the global network state $\vs\in\Ss$, where the entropy yields $H(\vo_1, \ldots, \vo_K)= H(\vs)$ and $H(\vo_k)<H(\vs)$ for $k\in\Ks$. Let $\vp:=[\vp_1, \ldots, \vp_K]$ and $\vo:=[\vo_1, \ldots, \vo_K]$ denotes the collections of HO control parameters and observations in the multi-cell network system respectively. The objective is to minimize the global cost function $\Js(\vs, \vp)$ with the information contained in the observations $\vo$. It is obvious that to learn a centralized model that takes all actions and observations $[\vp; \vo]\in\R^{(NM)^K}$ into account will lead to an extremely high model complexity. 

\begin{figure}[t]
      \centering
          \includegraphics[width=\columnwidth]{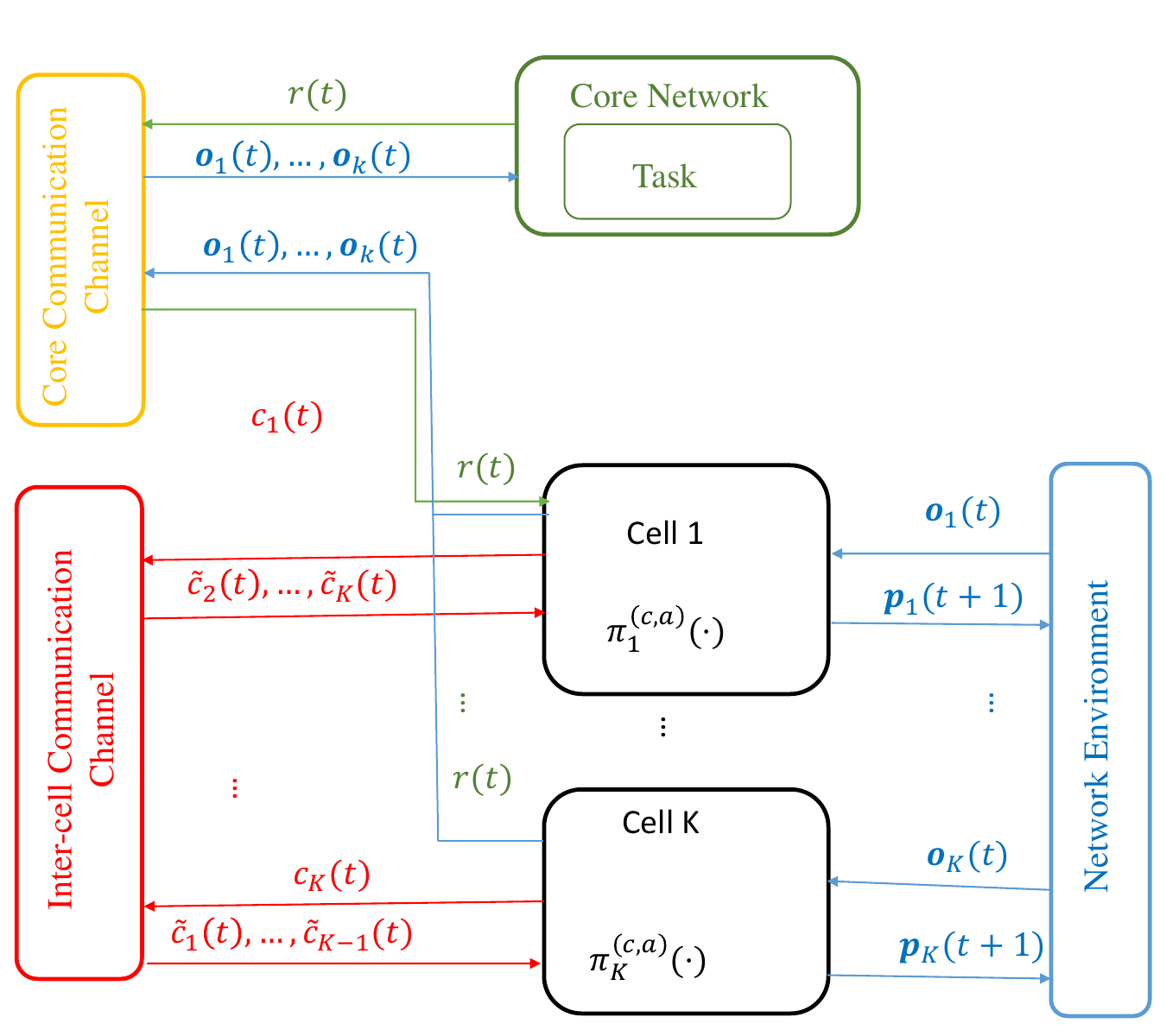}
      \caption{Task-oriented communication design for mobility load balancing problem in 5G and beyond SON}
      \label{fig:SON_Arch}
\end{figure}
To reduce the computational complexity, a task-oriented communication design can be considered as shown in Fig. \ref{fig:SON_Arch}. Such design enables distributed execution of the joint training of the local communication and control polices $\pi_k^{(c, a)}$, $k\in\Ks$. At time slot $t$, each cell observes a partial information $\vo_k(t)$ from the environment, with the policy $\pi_k^{(c, a)}(t)$, it derives an encoded message $\vc_k(t)\in\Cs_k$ which extracts the information of the local observation, and sends it to other cells via inter-cell communication channel (e.g., the X2 interface in 5G networks). After receiving the messages from the other cells $\left\{\tilde{\vc}_l: l\in\Ks\setminus\{k\}\right\}$ (note that with lossless channel we may have $\tilde{\vc}_k = \vc_k$), the local policy $\pi_k^{(c, a)}(t)$ also decides the new configuration of the HO parameters $\vp_k(t+1)$. On the other hand, to evaluate the global performance and compute the reward for all cells, each cell sends its observation to the core network (we assume lossless channel between the cells and the core network since they usually communicate with wired connection), and the core network who defines the task computes the common reward $r(t)$ based on the received observations, and sends it to the cells as the feedback of the current state $\vs(t)$ and joint actions $\vp(t)$. Each cell learns jointly a policy $\pi_k^{(c, a)}(t):\R^M\times\Pi_{l\in\Ks\setminus\{k\}}\Cs_l\to \Cs_k\times\R^N$ that maps the local observation $\vo_k$ and received messages $\{\vc_l:l\in\Ks\setminus\{k\}\}$ to a tuple of the local encoded message $\vc_k$ and local control parameters $\vp_k$. In this way, the model in each cell has an input space with the cardinality $|R|^M\prod_{l\in\Ks\setminus\{k\}}|\Cs_l|$, which is dramatically smaller than the input space cardinality $|R|^{(NM)^K}$ of the aforementioned centralized training based on the full observations and actions $(\vo, \vp)$.

\subsection{Tactile Internet} \label{subsection: applications - tactile}

\begin{figure}[t]
	\centering
	\includegraphics[width=1.03\columnwidth]{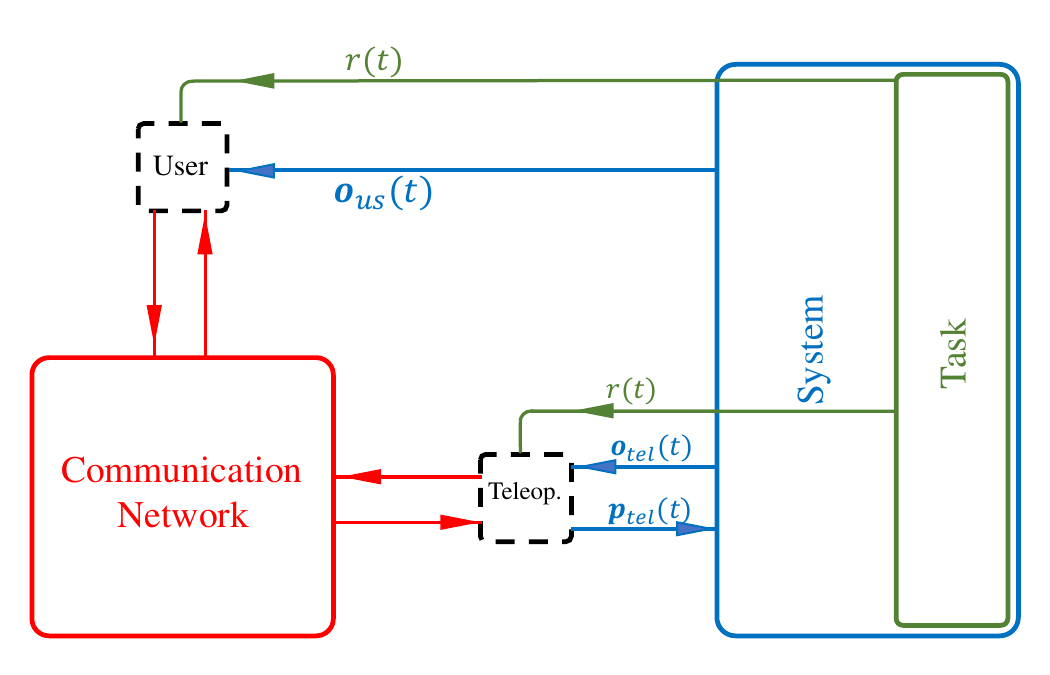}
	\caption{TOCD framework applied to the Tactile Internet.  	}
	\label{fig:TactileInternet}
\end{figure}

To enhance the degree of immersion of the user in distant communications, it is known that the communication of haptic information can play a crucial role. In the well-known scenario of teleoperation/telepresence, a human user interacts with a remote environment through: (i) a human system interface, (ii) a communication link, and (iii) the teleoperator. Such interactions involve both communications messages and action policies and should be carefully designed. Fig.~\ref{fig:TactileInternet} presents how the proposed task-oriented communication framework can be used to model Tactile Internet scenario. 

When teleoperation is performed over a communication channel with potential delays, noise, and uncertainties, it can be shown that achieving good performance in the teleoperation task can be formulated as a task-oriented communication problem \cite{shree20tactile}.
Haptic information can be divided in two different classes. The first class, called kinesthetic information, includes data related to muscle activation and movements. The second class, called tactile data, refers to the perception of pressure, texture, and temperature \cite{8399482}.

Communication of the first category of haptic data, kinesthetic data, involves the movement of an actuator of the teleoperator in such a way that a particular task is done with the best possible performance. Some examples for the tasks that require communication of kinesthetic data can be medical teleoperations or playing a musical instrument remotely. In the both of the mentioned examples, achieving a good level of performance in the task is not equivalent to reducing the distortion between a reference (desired) action signal and the controlled action process. While a wrong movement of teleoperator parallel to the surface of a piano keys generates no undesired musical note, an error in the controlled action process along the axis vertical to the surface of the piano can generate a wrong musical note. Let us recall that even wrong movements vertical to surface of the piano, depending on the location of the actuator and the magnitude of actuation error, may or may not cause any cost in task. Accordingly, the cost function of the task cannot be simply characterized by a mean squared error of the action process. Instead, a task-oriented cost function should be considered to achieve (close to optimal) performance in the task. Moreover, communication of kinesthetic data is of one particular sensitivity. Due to the stability requirements of the control loop on the teleoperator end, signal processing algorithms that are used at both sending and receiving ends should not cause large algorithmic delays \cite{5670635,otanez2002using}.

On the other hand, task-oriented communication of tactile information, the second category of haptic information, can also contribute to the improvement of the system performance. In this case, the cost function of the task is imposed by the perceptive abilities of an average human. Although there might be a difference between the communicated tactile signal and the signal received at the other end of the communication channel, if the difference between the two signal is beyond the perception of an average human, no cost should be associated to this communication error \cite{shahabi2002comparison, 4374151}. The authors in \cite{4374151} utilize the deadband principle to reduce the rate of communication of haptic information. The deadband principle is understood based on the perceptive abilities of an average human. According to this principle, the haptic information is transmitted only if the difference between the last communicated haptic data and the current available haptic data is perceptible by the human operator. While it is obvious that widening the deadband will result in more distortion, this may not affect the performance of task, which can be measured by the level of preciseness the task is performed by the user \cite{5670635}. Accordingly, finding the relationship between the achieved compression rate of the haptic information and the performance of the task is considered to be of substantial value \cite{5670635}.

The above-mentioned relationship is studied in \cite{kostina2019rate}, under several assumptions: (i) the uplink channel is noise-free, (ii) the cost function is considered to be a regularized quadratic function of the error in state of the system, (iii) the downlink channel noise is an additive white Gaussian noise, and (iv) and the system state is generated by a stochastic linear model.


    
    
     
    
%
%
%
%

%
%
%
\section{Challenges and Open Problems}\label{sec:open problems}
In this section, we envision challenges in developing efficient task-oriented communications solutions in future cyber-physical systems. After that, we present various potential research problems based on the task-oriented design framework.
We enumerate the existing challenges in two different categories, fundamental challenges of the framework and application specific challenges.

\subsection{Challenges of the Framework}

\subsubsection{Reward Signals}
The current framework assumes the availability of the common reward at all agents at no cost. In practice, the reward has to be sent to the agents via (wireless) communication channels. Therefore, the distribution of rewards among the agents must be taken into account when designing the communication strategy in such cases. Another challenge in designing task-oriented communications system is how to properly take into consideration different reward functions for different agents. In an extreme case, the agents can be competitive rather than collaborative \cite{Lowe2017:AC:MixedCoopComp}. In these cases, the communications should be adapted to specify individual target, which in turn affects both action and communication policies of the agents. 

\subsubsection{Training} Another issue is how the training is performed. It is well known that centralized training can provide optimal solution for the multi-agent system. However, the centralized training architecture is not always available, especially in dynamic multi-agent systems. This requires efficient distributed training design, in which the communications among the agents during the training phase is crucial and needs to be properly designed. So far, the communication is optimized based on the codebook and resource allocation philosophy. While this method is efficient for point-to-point and small networks, it is difficult to generalize to massive networks or taking latency and privacy aspects into account.

\subsubsection{Intrinsic Features of the Task} In task-oriented approaches, we are aiming to maximize system performance towards the task related KPIs given all the constraints on the communication resources, e.g., rate and latency. One of the unique aspects of the task-oriented communication problems that were absent in the “engineering problem\footnote{As mentioned by Shannon in his landmark work \cite{shannon1948mathematical}}”, is the remarkable impact that the characteristics of the task can have on the analytical studies. Oftentimes, the existing works on the task-oriented communication design, cannot predict/guarantee how their prescribed solutions would perform in general at every possible task. That is to say, before we apply a specific task-oriented scheme to a specific task, our analysis can hardly indicate how helpful the algorithm would be in improving effectiveness of using the communication resources - while achieving the task goals \cite{mostaani2020task,tung2021effective}. There are some intrinsic features in all tasks that would specify the extent to which a part of the original communication message to be transmitted can be discarded (see e.g., \cite{pappas2021goal}) and how they describe "semantics of information" metric to be a "context dependent" metric that maps qualitative information attributes to their "application dependent value". It is, however, very unclear yet what these intrinsic task features are and how they impact the extent to which we can improve the effectiveness of using communication resources. Finding our these features and how they influence the extent to which communication resources can be used effectively are of the unique theoretical challenges faced in this framework. 

\subsubsection{System Memory} While many of the existing solutions for task-oriented communication design are relying on the memorylessness of the source of information \cite{liu2021rate,kipnis2021rate}, this assumption is violated in almost every control task, where the current state of the system is determined based on its previous state(s) and the latest control decision made. Violating this assumption makes it hard (in general) to use classical tools offered by information theory that rely on asymptotic equipartition property property of the information processes.

\subsubsection{Temporal Dynamics} In complex topologies, the temporal dynamics of the communication network are an important challenge. The physical mobility of certain agents is a prime factor for the temporal dynamics and in combination with the excessive environment state space can lead to volatile behavior. In this context, communication diversity and connectivity prediction tools, directly dimensioned by their impact on the overall task effectiveness, should be able to produce stable/robust communication and action policies.

\subsubsection{Scalability} Although not a surprising challenge, scalability of the envisioned methodology is of paramount importance, especially in cases of large networks of agents with limited capabilities. Independently of the mathematical tool of choice (e.g. dynamic programming, optimization, learning), generalizing from a toy connectivity example to practical sizeable configurations will require unconventional approaches to the problem (e.g. large system approximations).





%
%
%
\subsection{Application-Specific Challenges}
\subsubsection{Industrial Internet-of-Things Networks}
The agents in IIoT systems differ from other applications in two main features: light computational capability and limited energy budget. Consequently, energy-efficient communications is expected to  be the first design criteria for IIoT applications. Meanwhile, IIoT systems usually demand timely decision-making actions and consequently stringent latency, which contradicts to the energy-efficiency target. This trade-off between energy efficiency and reliability asks for new design method for IIoT networks. URLLC is a promising solution for designing the communications in multi-agent IIoT systems. However, the current URLLC design does not take into account the requirements for the learning process. Therefore, the URLLC property should be jointly considered with the action policies considering the energy and computation capacities of IIoT nodes.  
\subsubsection{UAV Communications Networks}
UAV communications plays an essential role in the future communications networks as not only a stand-alone system in dedicated areas but also complementary parts of the cellular networks \cite{vinogradov2019tutorial}. More specifically, multiple UAVs can cooperate to provide communications in isolated areas for rescuing or sensing purposes, or they can aid the macro BSs to enhance handover or provide ultra-reliable communications to ground users \cite{azari2018ultra}. In either cases, reliable communications among the UAVs is key to improve the overall performance of the UAV networks. One challenge in UAV communications networks is how to jointly design the trajectory of every UAV to reduce collision risk and improve the system energy efficiency \cite{YuanTVT21}. However, since the UAVs usually operate in dynamic environment, conventional methods may not be applied due to the lack of proper system modeling. In fact, the UAVs' optimal trajectories are difficult, and sometimes unable to obtain as they depend on the movements and actions of all the UAVs. Therefore, one efficient way is that the UAV optimizes its trajectory and action policy while listening to the others, which requires efficient communications among the UAVs. Efficient communications should cover both how and what to communicate. On one hand, the UAVs should have to be well coordinated in accessing the common channel to avoid interference. On the other hand, each UAV has to determine what message to communicate to other UAVs in order to maximize the UAVs' collaboration. As a result, this asks for a novel design paradigm that optimizes the communications based on specific semantics. 
\subsubsection{Federated Learning}
Federated learning (FL) is an emerging distributed ML framework that allow a large number of edge nodes collaboratively train a shared learning model \cite{kairouz2019advances}. FL is capable of addressing many challenges in implementing ML over networks \cite{ParkProIEEE19,park2020communicationefficient}.
One of they key challenges in FL is non independently and identically distribute (non-i.i.d.) data among devices
and resource constraints \cite{StichICLR2019,ZhangJMLR13,WangJSAC2019}. Because the update at the edge nodes are based on their locally available data, the contribution to the aggregated system parameters varies from one edge node to another. As a result, always-transmit policy is no longer the optimal transmission in FL. In fact, a node which does not have sufficient data usually generates bias gradient parameters. Sending these parameters to the server does not improve, and sometimes harms the learning process. This asks for a context-aware transmission policy to tackle this issue. It is shown in \cite{Dinh20,DinhTWC22} that a proper transmission policy can improve the FL in terms of both energy efficiency and convergence performance. Another issue in FL is the different privacy levels among the edge nodes \cite{DuchiACM18,mcmahan2018learning}, which requires unequal-protection transmission and coding designs. Optimizing the source-channel coding to satisfy the privacy requirements and improve the learning convergence is usually difficult and requires novel system design perspectives.

\subsubsection{Mobile Edge Computing}
MEC will be a key component in the 5GB architecture to implement the intelligence on the network edge. By being equipped with both computational and storage capabilities, MEC nodes are able to determine to perform requested tasks locally or to offload them to the cloud server. Most of the existing works consider single MEC node that can optimally make task offloading decision. When applying to multiple cooperative MEC nodes, such methods, however, no longer render optimal policies. This is because in multiple MEC agents context, one node's action can have affects on other nodes. Furthermore, as the MEC nodes usually share the same communication medium, e.g., channel bandwidth, it requires proper transmission design to efficiently mitigate interference among the MEC agents. The major challenge is how to jointly design the MEC's local action policy with communications policy to balance the exploration-exploitation tradeoff. The authors of \cite{Cao2020} have shown the potential of such joint design for IIoT systems, in which edge-device acts as a machine-type agent (MTA). The MTAs collaboratively learn optimal policy for channel access and task offloading in multi sub-carrier D2D environment. At every time slot, each MTA determines to compute the task locally or to offload the task to the MEC server. By modeling the state space including offloading decisions, channel access status and computation task, the authors proposed to use multiagent deep deterministic policy gradients algorithm on actor-critic network at each MTA.  then the MTAs exchange their locally trained model to generalize the global model. The benefit of multiagent MEC system presented in \cite{Cao2020} is based on an assumption that all the edge nodes can perfectly communicate to each other during the training phase. In many practical cases, such assumption rarely occurs due to imperfect communications among the edge nodes and the highly dynamic network topology. This asks for novel distributed designs of action and communication policies. 

\subsubsection{5G and Beyond Self-Organizing Networks}\label{subsec:5GSON}
Although it has been almost a decade since the concept of SONs was introduced to the next generation mobile networks (NGMN) and 3GPP standards \cite{TS32500}, the existing SON solutions have not met the high expectation of the operators to achieve a fully self-aware cognitive network with automated configuration, monitoring, troubleshooting, and optimization. This is because, with the emergence of new wireless devices and applications, it is expected that a large amount of measurements and signaling overhead will be generated in future networks, while partial and inaccurate network knowledge, together with the increasing complexity of envisioned wireless networks, pose one of the biggest challenges for SON – maintaining global network information at the level of autonomous network elements is simply illusive in large-scale and highly dynamic wireless networks. Another big challenge is the network-wide optimization of strongly interdependent network elements, with the goal of improving the efficiency of total algorithmic machinery on the network level. Thus, to deliver SON solution for the 5G and beyond networks, we need to answer the following two questions:
\begin{enumerate}
    \item What information should be communicated among  network elements to enable  cooperation?
    \item How to let the network elements achieve consensus with a limited number of probes to improve performance on the global network level?
\end{enumerate}

Task-oriented communications system design can be leveraged to solve the above-mentioned two problems. Instead of considering the selection of the network state information and the optimization of the learning function as two independent processes, we can jointly optimize the information exaction and control optimization problems to improve both data efficiency and computational efficiency for large-scale highly interdependent network systems. 

\subsubsection{Autonomous Vehicles and Cooperative Automated Driving}
As highlighted previously in Sec. IV-E, the main tasks to be considered in the design of autonomous vehicles include pedestrian detection, gathering of 3D information about surrounding environment,  estimation of time to cross the street, action recognition, prediction, lane change decision and interaction of driver's interaction with the autonomous driving system. Furthermore, for cooperative driving systems, the important design tasks to be considered include cooperative communications among vehicles, infrastructure nodes and road-side units to avoid the collisions, cooperative platooning decisions under mobility and uncertainties and capturing situational awareness of the road-side information and required adaptation in dynamic situations are important design tasks to be considered. Furthermore, the positioning accuracy of autonomous vehicles may be impacted by issues like latency and packet loss in the underlying V2X communications networks; and in order to reduce the position errors and possible collisions, it is crucial to design suitable cooperative driving and merging strategies \cite{Xugroupingbaseddesign}. Another key issue to be addressed is to design robust and reliable cooperative sensing in order to effectively localize the surrounding and road-side objects detected by the onboard sensors and neighboring vehicles based on the available limited data-set \cite{HeCAVresearch2019}. Furthermore, DRL-enabled design of motion planning for autonomous vehicles, comprising trajectory planning, control and strategic decisions, is another promising area for future research, in which the main tasks to be designed include the environmental modeling, generating model abstractions, realization of underlying neural networks, and modeling of states, actions and rewards \cite{Aradi2020}. Furthermore, from the practical implementation and business perspective, it is interesting to investigate a task-based design in order to enable the cooperation among the key players of autonomous vehicles such as car makers, telecommunication industries and policy makers by balancing their individual interests.  

\subsubsection{Advanced RL Techniques for Task-oriented Communications}
Several advanced versions of RL including Inverse IRL, Safe RL and Multiagent-RL (MARL) seem promising to enable task-oriented communications design in the considered application domains, i.e., AV systems, multi-UAV networks and multi-sensory systems including TI. Among these, IRL determines a reward function to be optimized for a learning agent given either the set of measurements of the agent’s behavior or sensory inputs over time in various situations, and benefits from 
better inherent transferability of the reward function in new dynamic environmental situations as compared to the learned policy in the standard RL, which might need to be discarded in the changed situations \cite{Saurabh2021}. The main research issues associated with IRL along with some recommendations are included below: (i) existing works have mostly considered small-state spaces, which may not fully capture practical autonomous scenarios. To address this, one promising way would be to exploit the deeper version of IRL method by utilizing deep Q-network and other deep learning architectures \cite{Fernando2021}; (ii) learning a reward function is ambiguous as several reward functions may correspond to the same policy, resulting in the need of suitable accuracy metric to have the fair comparison of the agent’s behavior generated from the inferred reward function with the true expert’s behavior, and (iii) due to the involved iterative process comprising a constrained search over a space of reward functions, the complexity of the solution to the problem may grow disproportionately along with the problem size and number of required iterations, demanding for the need of low-complexity methods.

On the other hand, safe RL deals with finding suitable learning policies for the problems/applications (i.e., robotic systems, AVs), where it is essential to respect safety constraints and/or to guarantee reasonable system performance. As compared to the standard RL whose main objective is the long-term return maximization based on the real-valued reward, safe RL aims to consider both the long-term reward maximization and safety involved in the underlying agents/system as the first objective does not necessarily avoid the risk/negative outcomes due to inherent uncertainty of the underlying environment \cite{Garcia15a}. The main issues in employing safe RL along with some future recommendations are included below: (i) the employed aggressive exploration policy in model-free RL techniques to construct an accurate model might lead to high-risk situations, and also construction of a reasonably accurate model while capturing the underlying dynamics in a safe manner could be problematic in a model-based RL. To address these, some promising approaches could be learning dynamics from the demonstrations and employing policy and relational RL methods with bootstrapping \cite{Garcia15a}; (ii) most existing solutions are designed for finite MDPs, however, learning in realistic environments needs to deal with the continuous state and action spaces, thus leading to the need of devising suitable safe RL approaches, which can handle continuous actions and state space; (iii) selection of a risk metric might put limitations in using a particular RL algorithm, limiting the applicability of a safe RL algorithm to a particular application domain. This leads to the need of investigating generalized risk metrics and safe RL algorithms, which can be applied across different application areas. 

Similarly, MARL algorithms deal with the modeling of the multiple agents in the system, enabling coordination via information exchange among the agents and exploit the intention or hidden information of other agents' behaviors to have effective communications/cooperation. MARL techniques can find significant importance in various applications (i.e., AVs, cooperative cruise control, multi-UAVs) as they need to deal with the multiple learning agents and collaboration is essential among various learning agents in order to fully utilize the exploration space. Existing MARL techniques can be grouped under the frameworks of Markov/stochastic games and extensive-form games, and the former framework can be employed in three settings, namely cooperative (i.e., based on a common reward function or team-average reward), competitive (i.e., zero-sum Markov game) and mixed settings (general-sum game setting) \cite{Zhang2019multiagent}. The main issues associated with MARL along with some recommendations are highlighted below: (i) MARL algorithms may need to consider multi-dimensional goals, which can be often unclear, and they may fail to converge to the stationary Nash Equilibrium of general-sum Markov games. One approach to analyze the convergence behavior of MARL techniques is to utilize the concept of cyclic equilibrium; (ii) the stationarity assumption behind the convergence of single-agent RL methods becomes no longer valid for MARL methods, demanding for new mathematical tools for MARL analysis; (iii) joint state-action space needs to be taken into account by each agent, however, the dimension of this joint space can increase exponentially with the increase in the number of agents, leading to the issue of combinatorial MARL problem. One approach to tackle this scalability issue it to utilize deep neural networks to design MARL algorithms \cite{Nguyen2020deep}.

\balance

\section{Conclusion} \label{section: conclusion}
Task-oriented communication has been considered to be a new paradigm for designing communications  strategies  for  multi-agent  cyber-physical  systems. In this article, we have presented a comprehensive review and classification of the theoretical works across a wide range of research communities. We have then proposed a general conceptual framework for designing a task-oriented system and adapted it for the targeted use cases. Furthermore, we have provided a survey of relevant contributions in eight major application areas. Finally, we have discussed challenges and  open issues in the task-oriented communications design. 


\bibliographystyle{IEEEtran}
\bibliography{Bibfile.V4}


\begin{IEEEbiography}[{\includegraphics[width=1in,height=1.25in,clip,keepaspectratio]{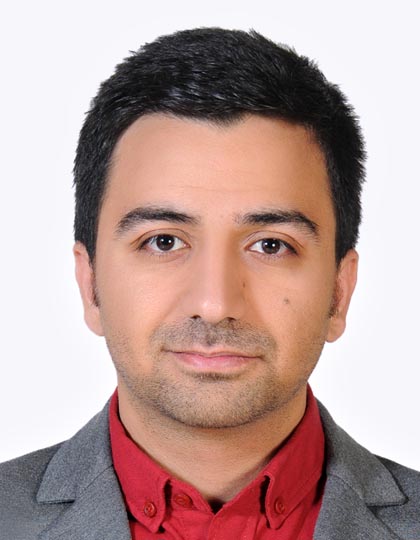}}]{Arsham Mostaani}
received his MS.C degree in electrical engineering from the University of Isfahan (Iran), in 2016. He has gained working experience as a researcher at the King's College London as well as Nokia Bell Labs - Stuttgart. His research interests include task-oriented communications, multiagent systems, wireless sensor networks. Arsham joined the Signal Processing and Satellite Communications group, SIGCOM, headed by Prof. Björn Ottersten and he will be advised by the latter.
\end{IEEEbiography}

\begin{IEEEbiography}[{\includegraphics[width=1in,height=1.25in,keepaspectratio]{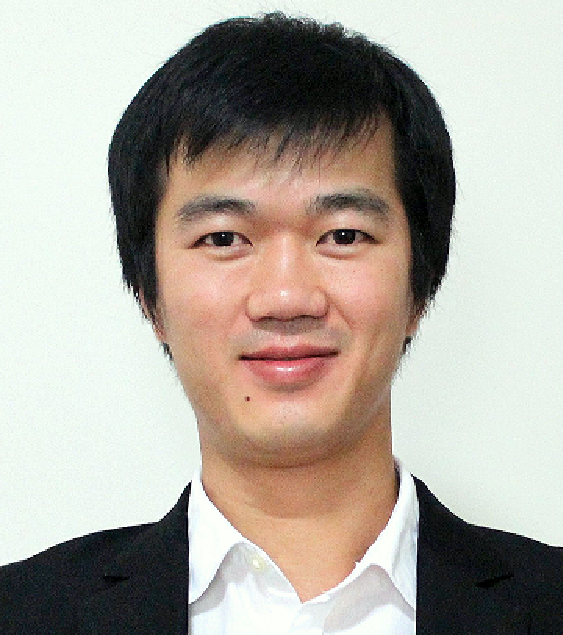}}]{Thang X. Vu} (Senior Member, IEEE) received the B.S. and the M.Sc., both in Electronics and Telecommunications Engineering, from the VNU University of Engineering and Technology, Vietnam, in 2007 and 2009, respectively, and the Ph.D. in Electrical Engineering from the University Paris-Sud, France, in 2014. In 2010, he received the Allocation de Recherche fellowship to study Ph.D. in France. 

From July 2014 to January 2016, he was a postdoctoral researcher with the Singapore University of Technology and Design (SUTD), Singapore. Currently, he is a research scientist at the Interdisciplinary Centre for Security, Reliability and Trust (SnT), University of Luxembourg. His research interests are in the field
of wireless communications, with particular interests of applications of optimization and machine learning on design and analyze the multi-layer 6G networks. He was a recipient of the SigTelCom 2019 best paper award. Currently, he is serving as an Associate Editor of the IEEE Communications Letters. He is a senior member of the IEEE.
\end{IEEEbiography}

\begin{IEEEbiography}[{\includegraphics[width=1in,height=1.25in,clip,keepaspectratio]{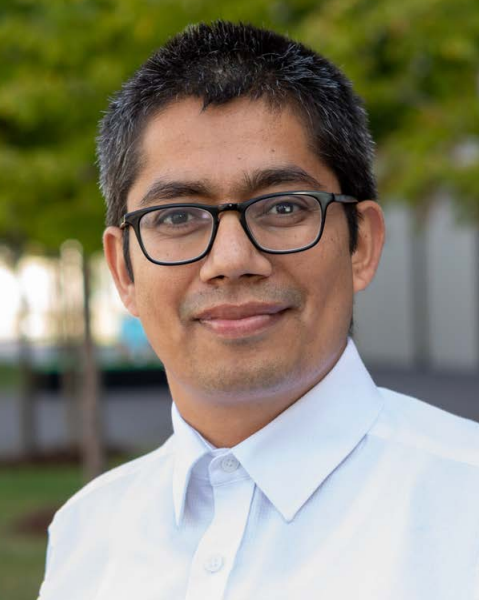}}]{Shree Krishna Sharma
}
(Senior Member,
IEEE) received the Ph.D. degree in wireless communications from the University of Luxembourg,
in 2014. He held various research and academic
positions at the SnT, University of Luxembourg;
Western University, Canada; and Ryerson University, Canada. He has published more than 100 technical papers in scholarly journals, international
conferences, and book chapters, and has over
4200 google scholar citations with an H-index of
30. He was a recipient of several prestigious awards, including ‘‘FNR Award
for Outstanding Ph.D. Thesis 2015’’ from FNR, Luxembourg, ‘‘Best Paper
Award’’ in CROWNCOM 2015 Conference, and ‘‘2018 EURASIP JWCN
Best Paper Award.’’ He was a co-recipient of the ‘‘FNR Award for Outstanding Scientific Publication 2019.’’ He is a Lead Editor of two IET books
Satellite Communications in the 5G Era and Communications Technologies
for Networked Smart Cities.
\end{IEEEbiography}

\begin{IEEEbiography}[{\includegraphics[width=1in,height=1.25in,clip,keepaspectratio]{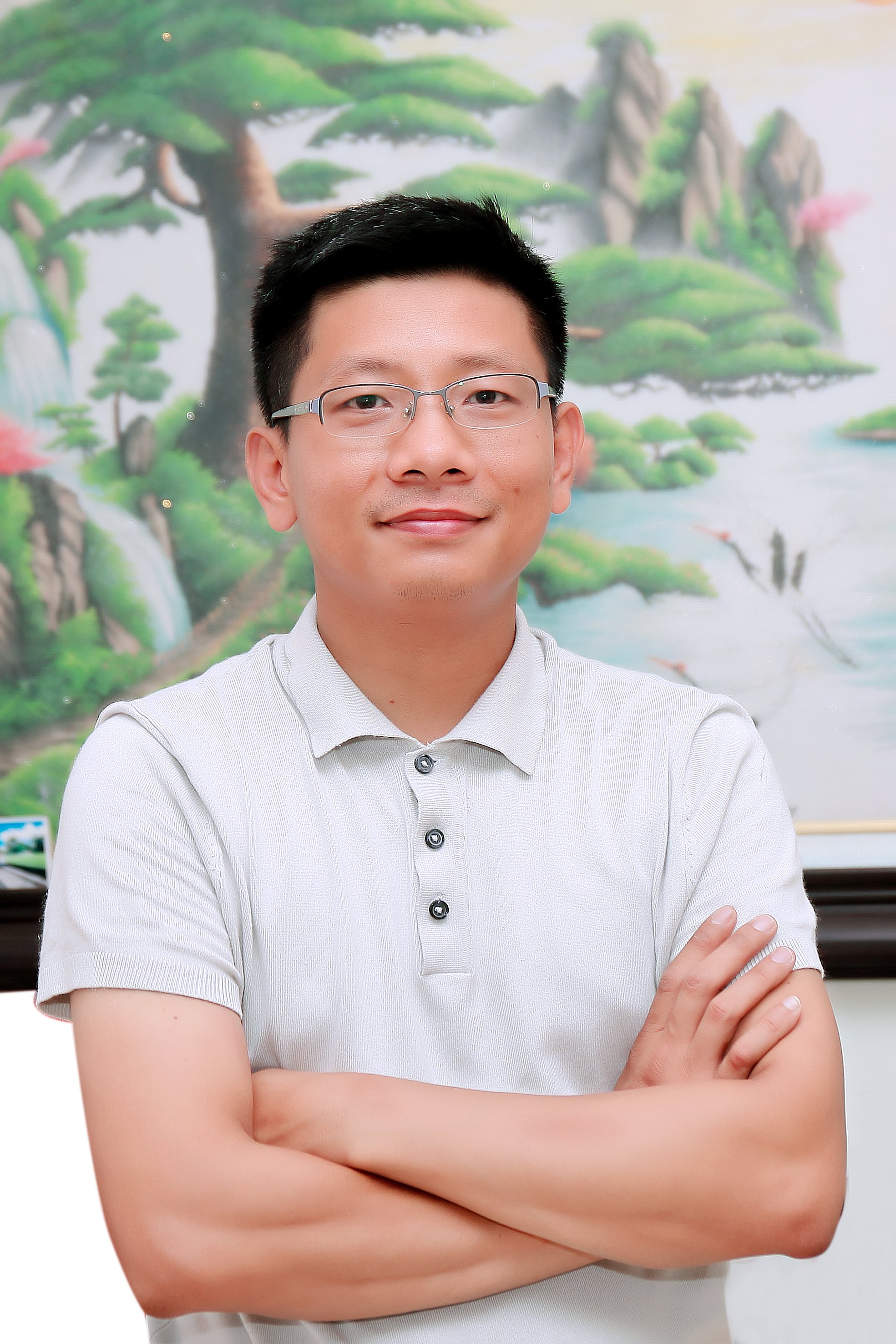}}]{Van-Dinh Nguyen }
(Senior Member, IEEE) received the B.E. degree in electrical engineering from Ho Chi Minh City University of Technology, Vietnam, in 2012 and the M.E. and Ph.D. degrees in electronic engineering from Soongsil University, Seoul, South Korea, in 2015 and 2018, respectively. Since 2022, he is an Assistant Professor with VinUniversity, Vietnam. He was a Research Associate with the SnT, University of Luxembourg, a Postdoc Researcher and a Lecturer with Soongsil University, a Postdoctoral Visiting Scholar with University of Technology Sydney,  and a Ph.D. Visiting Scholar with Queen’s University Belfast, U.K. His current research activity is focused on the mathematical modeling of 5G/6G cellular networks, edge/fog computing, and AI/ML solutions for wireless communications. 

Dr. Nguyen received several best conference paper awards, the Exemplary Editor Award of IEEE Communications Letters 2019, IEEE Transaction on Communications Exemplary Reviewer 2018 and IEEE GLOBECOM Student Travel Grant Award 2017. He has authored or coauthored in some 60 papers published in international journals and conference proceedings. He has served as a reviewer for many top-tier international journals on wireless communications, and has also been a Technical Programme Committee Member for several flag-ship international conferences in the related fields. He is an Editor for the IEEE Open Journal of The Communications Society and IEEE Communications Letters.

\end{IEEEbiography}

\begin{IEEEbiography}[{\includegraphics[width=1in,height=1.25in,clip,keepaspectratio]{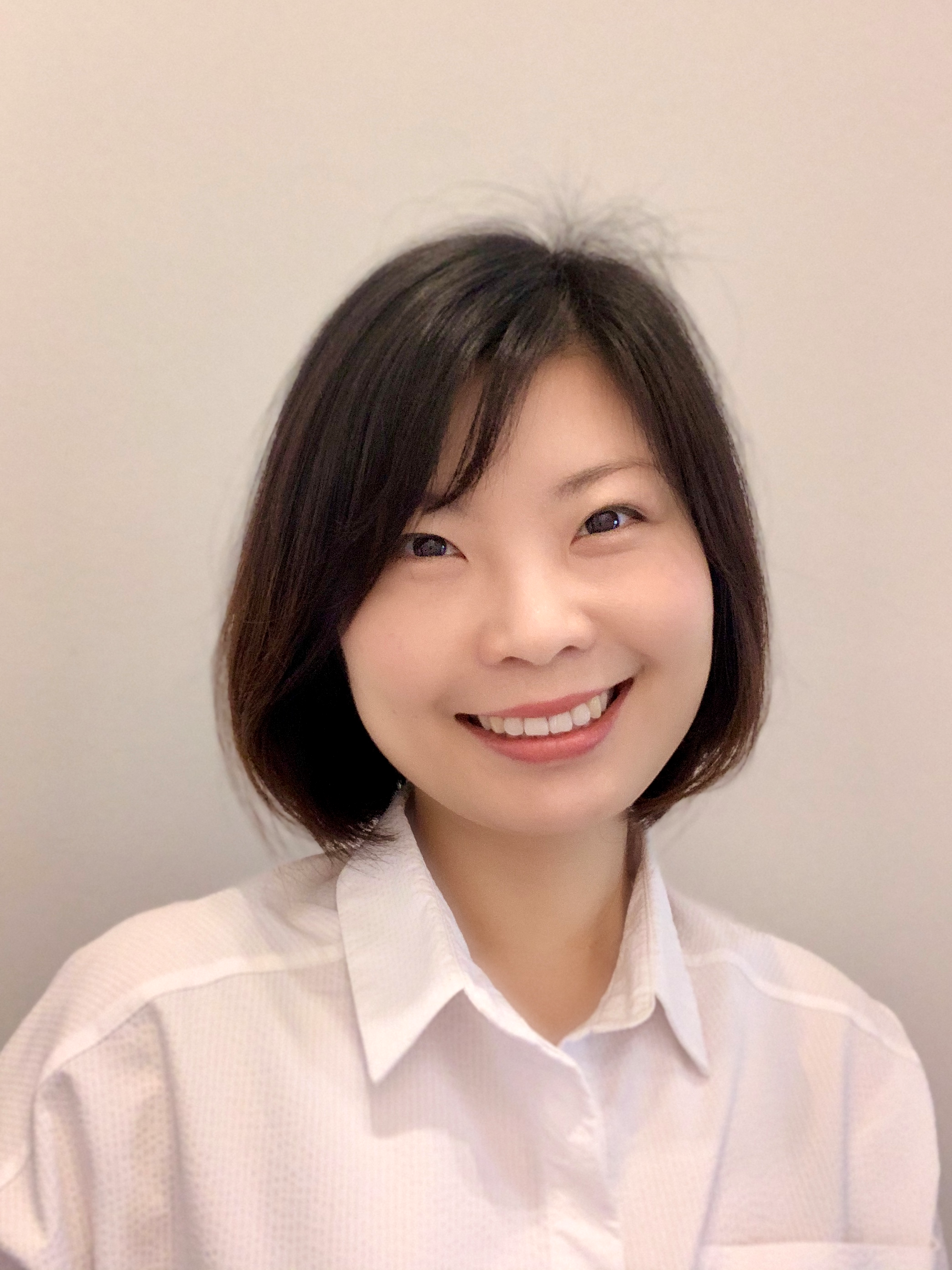}}]{Qi Liao}
Dr. -Ing. Qi Liao received the M.S. in E.E. degree and the Dr.-Ing. degree from Heinrich-Hertz-Chair for Information Theory and Theoretical Information Technology, Technical University of Berlin in 2010 and 2015, respectively. From 2010 to 2013 she was a research associate at Fraunhofer Institute for Telecommunications, Heinrich Hertz Institute, Berlin. From 2013 to 2014, she was a PhD intern in the Department of Statistics and Learning Research at Bell Labs in Murray Hill, US. Since 2015 she has been a research scientist at Nokia Bell Labs Core Research, Stuttgart, Germany. Her current research interests include optimization for multi-agent systems, resource allocation, stochastic optimization, and machine learning and deep learning techniques. She received the Best Paper Award from IEEE ICC in 2022. She holds more than 50 peer reviewed journal articles, conference papers, and granted or filed patents.
\end{IEEEbiography}

\begin{IEEEbiography}[{\includegraphics[width=1in,height=1.25in,clip,keepaspectratio]{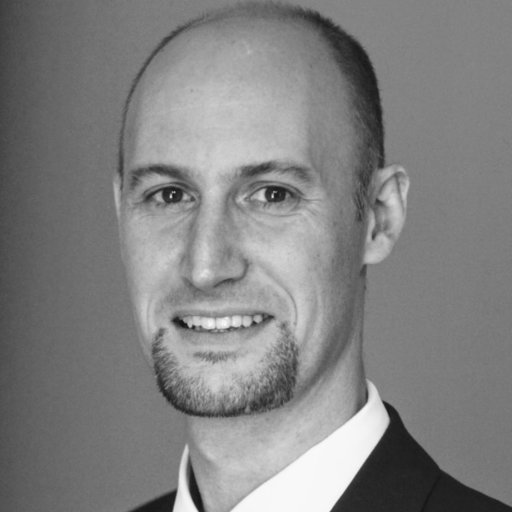}}]{Symeon Chatzinotas}
(Senior Member, IEEE) is currently Full Professor / Chief Scientist I and Head of the SIGCOM
Research Group at SnT, University of Luxembourg. In the past, he has been a Visiting Professor at the University of
Parma, Italy and he was involved in numerous Research and Development projects for the National Center for Scientific
Research Demokritos, the Center of Research and Technology Hellas and the Center of Communication Systems Research,
University of Surrey. He received the M.Eng. degree in telecommunications from the Aristotle University of Thessaloniki,
Thessaloniki, Greece, in 2003, and the M.Sc. and Ph.D. degrees in electronic engineering from the University of Surrey,
Surrey, U.K., in 2006 and 2009, respectively.
He was a co-recipient of the 2014 IEEE Distinguished Contributions to Satellite Communications Award, the
CROWNCOM 2015 Best Paper Award and the 2018 EURASIC JWCN Best Paper Award. He has (co-)authored more than
400 technical papers in refereed international journals, conferences and scientific books. He is currently in the editorial
board of the IEEE Open Journal of Vehicular Technology and the International Journal of Satellite Communications and Networking.
\end{IEEEbiography}

\end{document}